\documentclass[
aps,prb,twocolumn,letterpaper,twoside,
nobalancelastpage,raggedbottom,superscriptaddress,
amsmath,amssymb,floatfix,citeautoscript
]{revtex4-1}

\usepackage[T1]{fontenc}
\usepackage{times}

\usepackage{setspace}

\usepackage[pdftex]{graphicx}
\DeclareGraphicsExtensions{.png}

\usepackage{float} 

\usepackage{dcolumn} 
\newcolumntype{x}{D{.}{.}{1.5}}

\usepackage{color}
\definecolor{darkred}{rgb}{0.6,0.,0.}
\definecolor{darkgreen}{rgb}{0.,0.5,0.}
\definecolor{darkblue}{rgb}{0.,0.,0.6}

\usepackage[
breaklinks,
colorlinks=true,
linkcolor=darkred,
citecolor=darkgreen,
urlcolor=darkblue
]{hyperref}

\begin{document}

\def \bfa {{\bf a}}
\def \bfb {{\bf b}}
\def \bfA {{\bf A}}
\def \bfe {{\bf e}}
\def \bfd {{\bf d}}
\def \bfr {{\bf r}}
\def \bfq {{\bf q}}
\def \bfk {{\bf k}}
\def \bfR {{\bf R}}
\def \bft {{\bf t}}
\def \bfG {{\bf G}}
\def \calN {\mathcal{N}}
\def \olrho {\overline{\rho}}
\def \wtrho {\widetilde{\rho}}

\def \hatr {\widehat{\mathbf{r}}}
\def \hatx {\widehat{\mathbf{x}}}
\def \haty {\widehat{\mathbf{y}}}
\def \hatz {\widehat{\mathbf{z}}}
\def \hatn {\widehat{\mathbf{n}}}

\def \Real {\text{Re }}
\def \Imag {\text{Im }}
\def \Tr {\text{tr }}

\renewcommand{\figurename}{Figure}
\renewcommand{\tablename}{Table}

\newcommand{\OneColumnWidth}{8.5cm}
\newcommand{\OnePointFiveColumnWidth}{13.0cm}
\newcommand{\TwoColumnWidth}{18.0cm}
\newcommand{\CaptionLineSpacing}{1.2}

\title{Geometric stability of topological lattice phases}

\author{T. S. Jackson}
\email[]{tsjackson@physics.ucla.edu}
\affiliation{Department of Physics and Astronomy, University of California at Los Angeles, 475 Portola Plaza, Los Angeles, California 90095, USA}

\author{Gunnar M\"oller}
\affiliation{TCM Group, Cavendish Laboratory, J. J. Thomson Avenue, Cambridge CB3 0HE, UK}

\author{Rahul Roy}
\affiliation{Department of Physics and Astronomy, University of California at Los Angeles, 475 Portola Plaza, Los Angeles, California 90095, USA}

\date{\today}

\begin{abstract}
The fractional quantum Hall (FQH) effect illustrates the range of novel phenomena which can arise in a topologically ordered state in the presence of strong interactions. The possibility of realizing FQH-like phases in models with strong lattice effects has attracted intense interest as a more experimentally accessible venue for FQH phenomena which calls for more theoretical attention. Here we investigate the physical relevance of previously derived geometric conditions which quantify deviations from the Landau level physics of the FQHE. We conduct extensive numerical many-body simulations on several lattice models, obtaining new theoretical results in the process, and find remarkable correlation between these conditions and the many-body gap. These results indicate which physical factors are most relevant for the stability of FQH-like phases, a paradigm we refer to as the geometric stability hypothesis, and provide easily implementable guidelines for obtaining robust FQH-like phases in numerical or real-world experiments.
\end{abstract}

\maketitle

The fractional quantum Hall effect (FQHE) provides a spectacular manifestation of the breakdown of the spin-statistics relation in two dimensions: one obtains quantum number fractionalization\cite{Laughlin:1983hk,dePicciotto:1997dk,Saminadayar:1997dw,Camino:2005dq,Camino:2007ek} and, potentially, non-Abelian statistics\cite{mr} which can form the substrate for topologically robust quantum computing\cite{Nayak:2008dp}. Progress has been hampered by the considerable experimental difficulties involved in realizing the FQHE in the usual setting of a semiconductor heterostructure, but a flurry of interest in the field was set off by the recent insight\cite{Tang:2011by,Neupert:2011db,Sun:2011dk} that these exotic phases of matter may also arise in topologically nontrivial insulators with partially filled flat bands, or fractional Chern insulators\cite{Regnault:2011bu} (FCIs). The attractiveness of FCIs stems from the fact that the bandgap $\Delta$ may be set without the use of a large external magnetic field, the strength of which is one of the limiting factors in the semiconductor FQHE. There are currently a range of experimental proposals for realizing FCIs in cold atom systems\cite{Hafezi:2007gz,Cooper:2012bt,Cooper:2013jg,Yao:2012fn,Yao:2013eg}, transition metal oxides\cite{Xiao:2011eg,Venderbos:2012kz,Kourtis:2012hn} and elsewhere; a successful experimental implementation in ultracold fermions was recently announced in ref.~\onlinecite{Jotzu:2014vd}.

Moreover, FCIs raise theoretical questions independent of their experimental interest. The majority of theoretical work on the FQHE over its thirty-year history has focused on the influence of interactions on the Landau level Hamiltonian, which occupies a unique, highly symmetric point in the space of single-particle Hamiltonians. FCI phenomena constitute a non-trivial and poorly-understood generalization of the FQHE in which lattice effects are non-negligible; a generic FCI does not have the FQHE as a continuum limit, and examples of lattice effects without any continuum analog have already been noted\cite{Kol:1993hs,Moller:2009ir}. A generalization of our theoretical understanding of the FQHE to cover the case of FCIs is hence both nontrivial and experimentally relevant.

One possible approach to the stability of FCIs is via the single-mode approximation used by Girvin, MacDonald and Platzman\cite{Girvin:1985dm,Girvin:1986bu} (GMP), who made the ansatz that the most relevant excitations which destabilize an FQH ground state are neutral magnetoroton modes generated by the action of electron density operators projected to the lowest Landau level. These operators do not commute with each other, due to the projection, but GMP found that the set of operators remains a closed algebra under commutation. Intuitively, one expects that the form of this algebra plays a crucial role in the stability of the FQHE phases by limiting the set of possible destabilizing interactions. In a generic FCI, however, the analogous set of band-projected density operators is not a closed algebra, nor do the projected densities span the space of single-particle operators\cite{Parameswaran:2012cu}: there is no canonical mapping between a general lattice FCI and the continuum FQHE. 

In previous work, one of us\cite{Roy:2012vo} derived sufficient conditions for the band-projected density operators in an FCI to satisfy a closed algebra isomorphic to that present in the FQHE, which justified and elaborated upon a heuristic criterion used in previous FCI literature. Quantities describing the geometry of the Chern band (its embedding in Hilbert space) enter this analysis in a natural way as coefficients of terms which must necessarily vanish in order to obtain a closed algebra; remarkably, only three conditions need to be placed on the band's geometry for the isomorphism to be present to all orders in a long-wavelength expansion. Heuristically, one might expect that reproducing the density operator algebra would then suffice to reproduce the full physics of the FQHE, but this argument has not been fully tested in the literature.

In the present work, we report the results of extensive numerical simulations which demonstrate that quantitative measures based on the band-geometric conditions of ref.~\onlinecite{Roy:2012vo} are robustly correlated with the many-body gap in realizations of FQH-type phases in different FCI lattice models.
In addition to numerical data, we obtain several theoretical results, such as a scaling relation between the gap of an FQH-like state and the number of bands in an FCI model, which is essential for comparing different models. We find that the Berry curvature was computed incorrectly in a number of prior references; in Supplementary Note~1, we discuss why this quantity is defined unambiguously. 
The remarkably high degree of correlation we find between band geometry and the many-body gap leads us to propose a geometric stability hypothesis: that the algebra of band-projected density operators governs FQH-type phenomena in FCIs, even when the isomorphism doesn't hold exactly, and that the single-particle conditions investigated here are accurate qualitative estimators of the stability of an FQH-like state. This frames the theoretical problem of generalizing results on the FQHE to cover FCI physics by distilling the effects of the lattice into a small number of quantities measuring the relevant deviations of an FCI from lowest Landau level behavior. Our results are also of use in experimental design, as they provide a computationally inexpensive means to estimate which choices of FCI model parameters are most likely to yield a FQH-like state with the largest possible gap; a naive analysis of the scales involved has suggested this may be on the order of room temperature\cite{Tang:2011by}. From the opposite point of view, our results also indicate which areas of parameter space should be searched to find possible FCI states which do not correspond to FQH universality classes.

\section*{Results}

\subsection*{Geometry of Chern bands}

We begin by introducing the quantities studied below. A necessary ingredient in engineering a fractional Chern insulator is a flat, topologically non-trivial band, defined as follows. Let $|\bfR,b\rangle$ be a tight-binding orbital localized at position $\bfR+\bfd_b$; the Fourier transform of the $b$th basis orbital (where $b$ ranges from $1$ to $\calN$) is
\begin{equation}
\label{eq_tb_fourier}
|\bfk,b\rangle = \frac{1}{\sqrt{N_c}} \sum_{\bfR} e^{i\bfk \cdot(\bfR + \bfd_b)} |\bfR, b\rangle,
\end{equation}
where $\bfk$ is a crystal momentum restricted to the first Brillouin zone (BZ) and $N_c$ is the number of unit cells in the system, which are indexed by lattice vectors $\bfR$. Eigenstates of the tight-binding Hamiltonian are Bloch functions
\begin{equation}
\label{eq_efns}
| \bfk, \alpha \rangle = \sum_{b=1}^{\calN}  u^\alpha_b(\bfk) |\bfk, b \rangle,
\end{equation}
where $\alpha$ indexes the bands. At a fixed $\bfk$, the tight-binding Hamiltonian is an $\calN\times\calN$ matrix with entries 
\begin{equation}
\label{eq_band_ham}
H_{bc}(\bfk) = \sum_{\alpha=1}^{\calN} E_\alpha(\bfk) u_b^{\alpha \ast} (\bfk) u_c^\alpha (\bfk)
\end{equation}
and band energies $E_\alpha(\bfk)$. In general, for $\calN>1$, neither $H_{bc}(\bfk)$ nor $u^\alpha_b(\bfk)$ will have the periodicity of the reciprocal lattice, and this property is not needed to define the Berry curvature via eq.~(\ref{eq_curvature_def}) below. Imposing this periodicity by hand has led to demonstrably incorrect calculations in previous literature. We clarify this point with a discussion in Supplementary Note~1 and illustrate the consequences of incorrect computations in Supplementary Figs.~1,~2. 

Nontrivial topological order in a band $\alpha$ is indicated by a non-vanishing value of the (first) Chern number
\begin{equation}
c_1 = \frac{A_{BZ}}{2\pi} \langle B_\alpha \rangle,
\end{equation}
where $A_{BZ}$ is the area of the momentum-space Brillouin zone, $\langle\cdots\rangle$ denotes the average over the BZ, normalized so $\langle 1 \rangle=1$, and the Berry curvature\cite{Simon:1983du,Berry:1984ka} of the band $\alpha$ is defined as
\begin{equation}
\label{eq_curvature_def}
B_\alpha(\bfk) = -i \sum_{b=1}^{\calN} \left( \frac{\partial u_b^{\alpha \ast}}{\partial k_x}\frac{\partial u_b^{\alpha}}{\partial k_y} -  \frac{\partial u_b^{\alpha \ast}}{\partial k_y}\frac{\partial u_b^{\alpha}}{\partial k_x} \right).
\end{equation}
The leading-order condition (in terms of a long-wavelength expansion) for the existence of an isomorphism between the band-projected density operators and the GMP algebra found in ref.~\onlinecite{Roy:2012vo} is that the Berry curvature should be constant as a function of $\bfk$. In the results we report here, we quantify fluctuations of Berry curvature over the BZ  by their root-mean-square (RMS) value,
\begin{equation}
\label{eq_sigmac_def}
\sigma_B \equiv \sqrt{ \frac{A_{BZ}^2}{4\pi^2}\langle B^2 \rangle - c_1^2}.
\end{equation}
We normalize $\sigma_{B}$ in the same way as the Chern number, so that (\ref{eq_sigmac_def}) is dimensionless as well as insensitive to the scales over which deviations from the mean curvature occur. 
 
The higher-order conditions obtained ref.~\onlinecite{Roy:2012vo} involve the pull-back of the Fubini-Study metric on Hilbert space\cite{Provost:1980hs}, which we refer to below as the quantum metric. In terms of Bloch functions, it is given by
\begin{align}
\label{eq_metric_def}
g^\alpha_{\mu\nu}(\bfk) &= \frac{1}{2} \sum_{b=1}^{\calN} \left[ \left( \frac{\partial u_b^{\alpha \ast}}{\partial k_\mu}\frac{\partial u_b^{\alpha}}{\partial k_\nu} + \frac{\partial u_b^{\alpha \ast}}{\partial k_\nu}\frac{\partial u_b^{\alpha}}{\partial k_\mu} \right) \right. \\\
{}& \quad - \sum_{c=1}^{\calN} \left.\left(\frac{\partial u_b^{\alpha \ast}}{\partial k_\mu}u^\alpha_b u^{\alpha \ast}_c \frac{\partial u_c^{\alpha}}{\partial k_\nu} + \frac{\partial u_b^{\alpha \ast}}{\partial k_\nu}u^\alpha_b u^{\alpha \ast}_c \frac{\partial u_c^{\alpha}}{\partial k_\mu} \right) \right]. \nonumber
\end{align}
The next-to-leading order condition of ref.~\onlinecite{Roy:2012vo} is that the quantum metric also be constant over the BZ. We adopt
\begin{equation}
\label{eq_sigmag_def}
\sigma_g \equiv \sqrt{ \frac{1}{2} \sum_{\mu,\nu} \langle g_{\mu\nu} g_{\nu\mu} \rangle -  \langle g_{\mu\nu}  \rangle \langle g_{\nu\mu} \rangle }
\end{equation}
as the appropriate generalization of RMS fluctuation to tensor quantities. The final constraint on the band geometry is that
\begin{equation}
\label{eq-detineq}
D(\bfk) \equiv \det g^\alpha(\bfk) - \frac{B_\alpha(\bfk)^2}{4} = 0.
\end{equation}
It was shown in ref.~\onlinecite{Roy:2012vo} that the left-hand side of (\ref{eq-detineq}) is always nonnegative; the condition that it vanishes is equivalent to the condition that $g^\alpha$ and $F^\alpha = \left(\begin{smallmatrix} 0 & B_\alpha \\ -B_\alpha & 0 \end{smallmatrix}\right)$ are the real and imaginary components of a K\"ahler metric $h^\alpha = g^\alpha + i F^\alpha/2$. This means that, unlike the first two conditions, the metric determinant inequality $D(\bfk) \geq 0$ measures deviations from lowest Landau level physics, specifically. Analogous conditions may be derived for higher Landau levels. 

A stronger condition can be obtained by considering the trace of the quantum metric instead. It was additionally shown in ref.~\onlinecite{Roy:2012vo} that
\begin{equation} 
\label{eq-trineq}
T(\bfk) \equiv \Tr g^\alpha(\bfk) - |B_\alpha(\bfk)| \geq 0.
\end{equation}
In Supplementary Note~2 we show that if this inequality is saturated, the quantum metric is isotropic and $D(\bfk)$ must vanish. Hence the condition $T(\bfk)=0$ is equivalent to requiring that the algebra of band-projected density operators be identical to the GMP algebra, while $D(\bfk)=0$ merely requires that they be isomorphic.


\subsection*{Band geometry hypothesis}

The purpose of the present work is to investigate the degree to which the above criteria are satisfied in several FCI models known to exhibit FQH-like phases\cite{Wang:2011cy,Scaffidi:2012dx,Wu:2012do}. In this section, we outline procedures common to all models studied.

The stability of an FCI phase is trivially influenced by the dispersion of the occupied band. Fortunately, the dispersion of a band is independent of its Berry curvature and quantum metric: the former only depends on the Hamiltonian's spectrum while the latter depend only on its eigenvectors. This allows us to eliminate any dispersion-related confounding effects by energetically flattening the bands of each lattice model, which is equivalent to smearing nearest-neighbor hopping terms over an exponentially-localized area\cite{Neupert:2011db,Regnault:2011bu}.

Differences between Chern bands and Landau levels also enter in the form of the Hamiltonian's interaction term. Unlike the energetic considerations, this dependence is still poorly understood, so we have limited the scope of the present paper to on-site repulsive interactions only (which necessitates bosonic statistics), since this is the lattice interaction which most closely matches the isotropy present in the continuum. For each lattice model considered, we therefore investigate the bosonic Laughlin state at filling fraction $\nu=1/2$ (stabilized by a two-body delta-function interaction) and the bosonic Moore-Read state at $\nu=1$ (stabilized by a three-body delta-function interaction). These states have completely different topological orders; furthermore the Laughlin state is known to be more robust in general than the Moore-Read state, so examining both provides a useful probe of the sensitivity of band-geometric arguments. 

The band geometry hypothesis predicts that the most important factor will be Berry curvature fluctuations. Low curvature fluctuations were heuristically identified as a desirable criterion in the earliest FCI literature\cite{Tang:2011by,Sun:2011dk}, which has been well established by subsequent work (see in particular refs.~\onlinecite{Dobardzic:2013gc,Dobardzic:2014ei}). In the present work we therefore focus on the sub-leading conditions, namely the influence the quantum metric has on the gap. 

Fluctuations of the quantum metric are predicted to be the next most relevant quantity, but in the models examined this was found to have a high degree of linear correlation with the Berry curvature fluctuations (see Supplementary Fig.~3). There is no a priori reason this should be the case: other metric-derived quantities were found to be largely independent of Berry curvature. In addition, we found that the trace inequality (\ref{eq-trineq}) was far more correlated with the gap than the determinant inequality (\ref{eq-detineq}) for all models examined, despite corresponding to a stronger condition on the algebra of density operators. These findings go beyond the scheme laid out in ref.~\onlinecite{Roy:2012vo}. Due to space constraints, we present data on the dependence of the gap on the determinant condition for low values of $\sigma_{B}$ in Supplementary Fig.~4 for the kagom\'e lattice model and Supplementary Fig.~5 for the ruby lattice model.

\vspace{-1mm}
\subsection*{Haldane model}
\label{sec-Haldane}

\begin{widetext}

\begin{figure}[bth]
\centering
\includegraphics[width=\TwoColumnWidth]{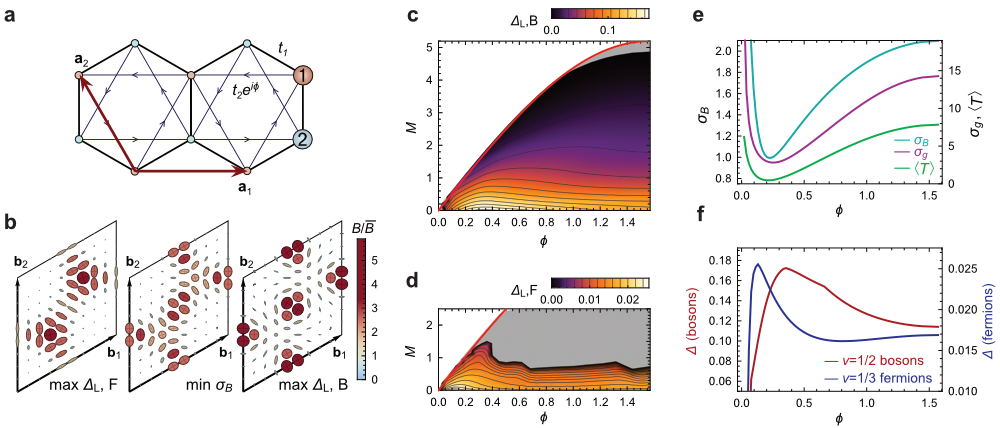}%
\caption{
\setstretch{\CaptionLineSpacing}
\textbf{Band geometry and gap data for the Haldane model.} 
(\textbf{a}) Honeycomb lattice used to define the Haldane model. Basis vectors $\bfa_{1}, \bfa_{2}$ are shown in red; basis elements are shown by differently colored/numbered sites. Hopping elements are shown with black and blue edges; arrowheads indicate the chirality convention for complex hoppings. (\textbf{b}) Band geometry over the reciprocal lattice unit cell spanned by $\bfb_{1}, \bfb_{2}$, for parameter values maximizing the gap and minimizing $\sigma_B$. Axes of ellipses are proportional to the eigenvectors of the quantum metric $g^\alpha (\bfk)$, and ellipse color is given by the relative deviation of Berry curvature $B_\alpha(\bfk)$ from its Brillouin zone-averaged value. (\textbf{c}) Gap $\Delta$ as a function of $(\phi, M)$ for $N=8$ bosons at $\nu = 1/2$ with an on-site repulsion. (\textbf{d}) Gap as a function of $(\phi, M)$ for $N=8$ fermions at $\nu = 1/3$ with nearest-neighbor repulsion. Note that this plot differs from Fig.~8 of ref.~\onlinecite{Wu:2012do} because we exclude (light gray) parameters failing to meet energetic and entanglement-based criteria for Laughlin-type order. (\textbf{e}) Berry curvature fluctuations $\sigma_B$ (left scale) and metric fluctuations $\sigma_g$ and average trace inequality $\langle T \rangle$ (right scale) as functions of $\phi$ at $M=0$. (\textbf{f}) Reproduction of the gap data from panels (\textbf{c}), (\textbf{d}) along $M=0$.
}
\label{fig-haldane}
\end{figure}

\end{widetext}

The first Chern insulator model was introduced by Haldane\cite{Haldane:1988gh}, who considered a tight-binding model on the honeycomb lattice with nearest- and next-nearest-neighbor hoppings (Fig.~\ref{fig-haldane}a) and a Peierls phase due to non-uniform threading of magnetic flux through each hexagon. The single-particle Hamiltonian for the Haldane model is
\begin{align}
\label{eq-h-haldane}
H_{\text{H}}(\bfk) &= t_1 \sigma_1 \sum_{i=1}^3 \cos \tfrac{1}{3}(k_i+2k_{i+1}) \nonumber \\
{} & \quad - t_1 \sigma_2 \sum_{i=1}^3 \sin \tfrac{1}{3}(k_i+2k_{i+1}) \nonumber \\
{} & \quad + \sigma_3 \Bigl( M-2t_2 \sin \phi \sum_{i=1}^3 \! \sin k_i  \Bigr).
\end{align}
where the Pauli sigma matrices act on the band index, $k_i\equiv\bfk\cdot\bfa_i$, $\bfa_3=-\bfa_1-\bfa_2$ and  $i=1,2,3$ is interpreted cyclically mod 3. The lower band has a nonzero Chern number when $|M/t_2| \leq 3\sqrt{3}|\sin\phi|$. This model has been extensively studied numerically, and the addition of short-ranged repulsive interactions has been shown to yield both the bosonic\cite{Wang:2011cy,Scaffidi:2012dx} and fermionic\cite{Wu:2012do} Laughlin states at appropriate filling fractions.

For purposes of comparison with refs.~\onlinecite{Wu:2012do} and \onlinecite{Dobardzic:2013gc}, we consider the model at $t_1=t_2=1$. The energy spectrum has band crossings for these parameters, meaning that the bands cannot be flattened by local operators and the model analyzed in those references is not adiabatically connected to (\ref{eq-h-haldane}). In practice, however, one is most interested in the $M=0$ subspace; the Hamiltonian then depends only on the combination $t_2\sin\phi/t_1$, and an increase in $t_2/t_1$ which removes the crossing may then be compensated by a shift in $\phi$ which leaves the Hamiltonian (\ref{eq-h-haldane}) unchanged up to a scale.

The momentum dependence of the Berry curvature and quantum metric is shown in Fig.~\ref{fig-haldane}b for parameters which minimize $\sigma_{B}$ and maximize the gaps for the Laughlin state of bosons and fermions; these values are listed in Supplementary Table~1. We see that the distribution of Berry curvature minimizing $\sigma_{B}$ interpolates between that which maximizes the gap for the bosonic and fermionic Laughlin states, and likewise the value of $\phi$ minimizing $\sigma_{B}$ lies between the values minimizing the gaps (Fig.~\ref{fig-haldane}e,~f). The band geometry argument doesn't distinguish the statistics of the underlying particles, due to the fact that the projected density operators are bilinear in particle operators and bosonic in either case. The $(\phi,M)$ parameter space may be sampled exhaustively; many-body gaps for the bosonic and fermionic Laughlin states are shown over the full topologically non-trivial region of parameter space in Figs.~\ref{fig-haldane}c,~\ref{fig-haldane}d. The largest gaps and most uniform band geometry both occur for $M=0$. 

The other band-geometric criteria are highly correlated with the curvature fluctuation $\sigma_{B}$ and yield little new information for this model (Fig.~\ref{fig-haldane}e), beyond being close to the location of the maximum gaps (Fig.~\ref{fig-haldane}f). We prove in Supplementary Note~3 that the remaining band-geometric criterion, the determinant condition (\ref{eq-detineq}), is necessarily saturated for any two-band model, but the trace condition (\ref{eq-trineq}) remains nontrivial here.


\subsection*{Augmented Haldane model}

Although the Haldane model at fractional filling exhibits a robust Laughlin state, its Berry curvature remains highly nonuniform even in the best case (Fig.~\ref{fig-haldane}b). We would like to be able to compare this model with the kagom\'e and ruby lattice models, in which more uniform curvature may be achieved. In addition, since we're interested in the sub-leading effects the quantum metric has on the gap, we want to examine band configurations for which the metric is more independent of the Berry curvature than in Fig.~\ref{fig-haldane}e.

This may be accomplished by adding a third-nearest-neighbor hopping term, with independent coupling $t_3$, to the Haldane model Hamiltonian (\ref{eq-h-haldane}). In the previous section, we saw that a sub-lattice chemical potential $M$ always reduces the gap of an FCI phase, so we set $M=0$ below. The remaining couplings in the new Hamiltonian may be parameterized by $t_2 \sin \phi / t_1$ and $t_3/t_1$, so the interesting region of parameter space is still two-dimensional. 

\begin{figure}[t]
\centering
\includegraphics[width=\OneColumnWidth]{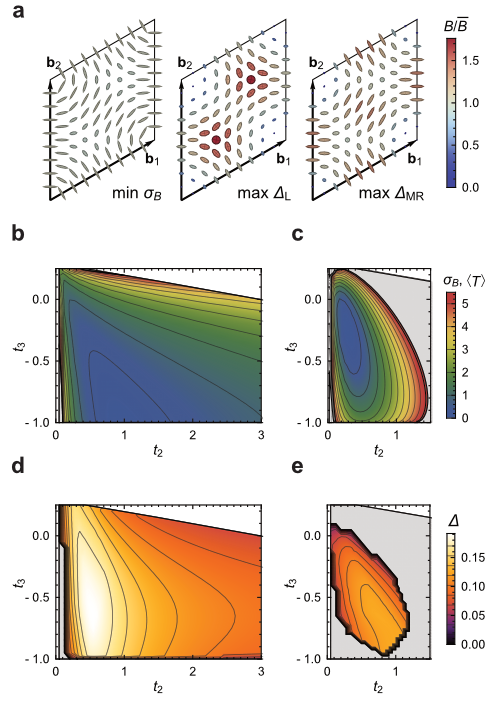}%
\caption{
\setstretch{\CaptionLineSpacing}
\textbf{Band geometry and gap data for the augmented Haldane model.} 
(\textbf{a}) Band geometry over the reciprocal lattice unit cell spanned by $\bfb_{1}, \bfb_{2}$, for parameter values maximizing the gaps and minimizing $\sigma_B$. Axes of ellipses are proportional to the eigenvectors of the quantum metric $g^\alpha(\bfk)$. Ellipse color is given by the relative deviation of Berry curvature $B_\alpha(\bfk)$ from its Brillouin zone-averaged value. (\textbf{b}) Berry curvature fluctuations $\sigma_B$ and (\textbf{c}) Brillouin zone-averaged trace inequality $\langle T \rangle$ as a function of couplings $t_{2}, t_{3}$, where we set $t_{1}=1$ and $\phi=\pi/2$ in this and remaining panels without loss of generality. (\textbf{d}) Gap $\Delta$ as a function of couplings for the bosonic Laughlin state of $N=8$ bosons at $\nu = 1/2$. (\textbf{e}) Gap as a function of couplings for the bosonic Moore-Read state of $N=10$ bosons at $\nu = 1$.
}
\label{fig-haldane-t3}
\end{figure}

The momentum dependence of the Berry curvature and quantum metric is shown in Fig.~\ref{fig-haldane-t3}a for parameters which minimize $\sigma_{B}$ and maximize the gaps for the Laughlin and Moore-Read states; these values are given in Supplementary Table~2. Comparison with Fig.~\ref{fig-haldane}b shows that curvature fluctuations have been reduced; furthermore, the minimum of $\sigma_{B}$ (shown in Fig.~\ref{fig-haldane-t3}b) occurs at different parameter values than the minimum of the trace condition (shown in Fig.~\ref{fig-haldane-t3}c). Gaps for the Laughlin and Moore-Read states are shown in Figs.~\ref{fig-haldane-t3}d and \ref{fig-haldane-t3}e, respectively; the maximum gaps in both cases occur at lower values of $t_2$ than the minimum value of $\sigma_B$, which one may attribute to the influence of $\langle T\rangle$.

\clearpage

\subsection*{Kagom\'e lattice model}
\label{sec-kagome}

\begin{figure}[bh]
\centering
\includegraphics[width=\OneColumnWidth]{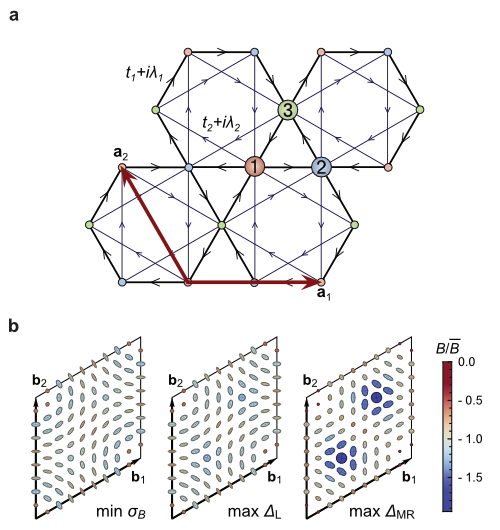}%
\caption{
\setstretch{\CaptionLineSpacing}
\textbf{Definition of couplings and band geometry for the kagom\'e lattice model.} 
(\textbf{a}) Lattice and chirality conventions for hopping terms in the kagom\'e lattice model Hamiltonian. Basis vectors $\bfa_{1}, \bfa_{2}$ are shown in red; basis elements are shown by differently colored/numbered sites. Hopping elements are shown with black and blue edges; arrowheads indicate the chirality convention for complex hoppings. (\textbf{b}) Band geometry over the reciprocal lattice unit cell spanned by $\bfb_{1}, \bfb_{2}$, for parameter values minimizing $\sigma_B$ and maximizing gaps, respectively. Axes of ellipses are proportional to the eigenvectors of the quantum metric $g^\alpha (\bfk)$. Ellipse color is given by the relative deviation of Berry curvature $B_\alpha(\bfk)$ from its Brillouin zone-averaged value.
}
\label{fig-kagome}
\end{figure}

A Chern insulator defined on the kagom\'e lattice was introduced by Tang, Mei and Wen \cite{Tang:2011by} (Fig.~\ref{fig-kagome}a). This model is attractive for our purposes since it has three bands, while remaining structurally similar to the Haldane model. 

Defining a complex hopping matrix element for the relative embedding of the sublattices in the unit cell as
\begin{equation}
\label{eq_hopping_def}
h_{bc}(\bfk) \equiv e^{i\bfk\cdot(\bfd_b - \bfd_c)} \widehat{e}_{b,c},
\end{equation}

where $\widehat{e}_{b,c}$ is the unit matrix whose $(b,c)$th entry is equal to $1$, the momentum space Hamiltonian for the kagom\'e lattice model is
\begin{align}
\label{eq_kag_ham}
H_\text{K}(\bfk) &= - \sum_{j=1}^3 \left[(t_1 - i \lambda_1) (1+e^{i\bfk\cdot\bfa_j}) h_{j,j+1} (\bfk) \right. \nonumber \\
{} &\qquad +  \left. (t_2 - i \lambda_2)(e^{i\bfk \cdot \bfa_j} +e^{i\bfk\cdot\bfa_{j+1}})h_{j,j+2} (\bfk) \right] \nonumber \\
{}&\qquad  + \text{h.c. }
\end{align}
where $\text{h.c.}$ is an abbreviation for the hermitian conjugate, $\bfa_3=-\bfa_1-\bfa_2$, and $j$ is interpreted cyclically mod 3. The relative offsets $\bfd_b$ are as depicted by the numbered sites in Fig.~\ref{fig-kagome}a. The momentum dependence of the Berry curvature and quantum metric is shown in Fig.~\ref{fig-kagome}b for parameters minimizing $\sigma_{B}$ and maximizing the gaps for the Laughlin and Moore-Read states; these values are listed in Supplementary Table~3.

\begin{widetext}

\begin{figure}[th]
\centering
\includegraphics[width=\TwoColumnWidth]{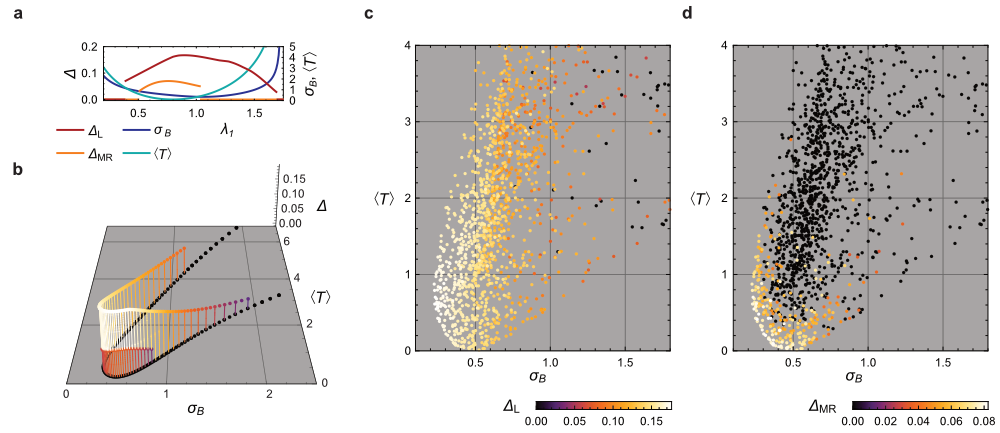}%
\caption{
\setstretch{\CaptionLineSpacing}
\textbf{Gap vs. band geometry for the kagom\'e lattice model.}  (\textbf{a}). Gap and geometry data for the kagom\'e lattice model with NN-only hoppings, as a function of the only coupling ratio $\lambda_{1}/t_{1}$. (\textbf{b}) The same data as a one-dimensional submanifold in band geometry space, which we parameterize in terms of Berry curvature fluctuations $\sigma_{B}$ and the averaged trace condition $\langle T \rangle$. The upper set (larger gap $\Delta$) of points are gaps for the $N=8$ bosonic Laughlin state at $\nu = 1/2$, while the lower are for the bosonic Moore-Read state at $\nu = 1$. (\textbf{c}) Gap for the kagom\'e lattice model with both nearest-neighbor and next-nearest-neighbor couplings, as a function of band-geometric parameters $(\sigma_B, \langle T \rangle )$ for the Laughlin state of $N=8$ bosons at $\nu = 1/2$.  (\textbf{d}) Gap as a function of $(\sigma_B, \langle T \rangle )$ for the Moore-Read state of $N=10$ bosons at $\nu = 1$, for the same coupling values.
}
\label{fig-kagome-gaps}
\end{figure}

\vspace{-5mm}

\begin{figure}[H]
\centering
\includegraphics[width=\TwoColumnWidth]{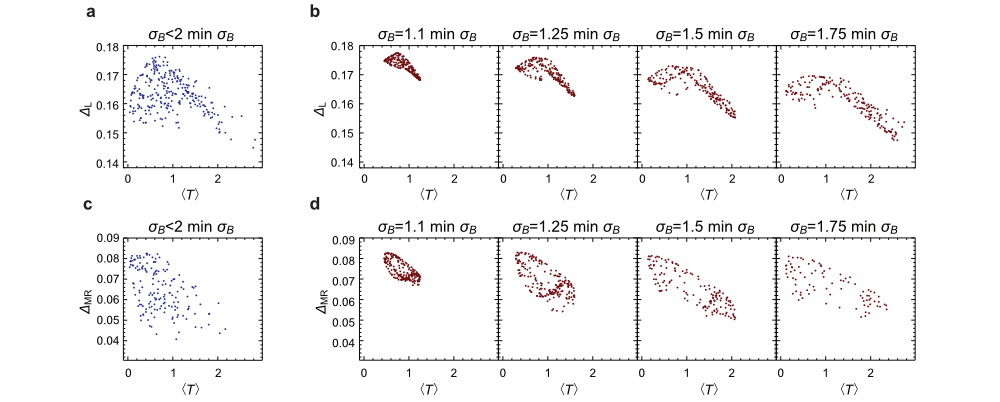}%
\caption{
\setstretch{\CaptionLineSpacing}
\textbf{Gap vs. trace condition for the kagom\'e lattice model subject to constraints on $\sigma_{B}$.}  (\textbf{a}) Gaps for the bosonic Laughlin state of $N=8$ bosons at $\nu = 1/2$, as a function of the Brillouin zone average of the trace condition $\langle T \rangle$. We only plot gaps for parameter values which have Berry curvature fluctuations $\sigma_{B}$ less than twice its minimum value. (\textbf{b}) Gap of the bosonic Laughlin state vs. $\langle T \rangle$ for parameter values randomly chosen on isosurfaces of constant $\sigma_B$ in the space of couplings. The parameter space sampling procedure used to obtain these sets of points is described in the Methods. (\textbf{c}),~(\textbf{d}) The same, for the bosonic Moore-Read state of $N=10$ bosons at $\nu = 1$. Note that the same sets of model parameters are used in each column.
}
\label{fig-kagome-shells} 
\end{figure}

\end{widetext}

\phantom{}\clearpage

We first consider the model with nearest-neighbor hoppings only (Figs.~\ref{fig-kagome-gaps}a,~b). Because band geometry is determined by single-particle quantities, the values of $(\sigma_B,\langle T\rangle)$ for a given value of the effective nearest-neighbor (NN) coupling $\lambda_{1}/t_{1}$ are identical for the Laughlin and Moore-Read states. Despite having different topological orders, gaps for both states decline monotonically as one proceeds from the region of minimum $\sigma_B$ and $\langle T\rangle$, and this trend holds over the entire phase (i.e., up to the closure of the gap.)

To establish that the above trends were not coincidental, we studied the entire $c_1=-1$ phase containing the NN-only point $t_1=\lambda_1$; $t_2=\lambda_2=0$ for the bosonic Laughlin (Fig.~\ref{fig-kagome-gaps}c) and Moore-Read (Fig.~~\ref{fig-kagome-gaps}d) states. The gap's sensitivity to band geometry is most apparent for the more fragile Moore-Read state: the state is only stable in a small region, with the largest gaps (white points) attained at parameters with the lowest values of $\sigma_{B}$ and $\langle T\rangle$ found. The fact that the region of stability is an arc, rather than a vertical line, demonstrates that $\langle T\rangle$ describes independent, non-negligible factors influencing the stability of this state. These phenomena are less evident in the Laughlin state (Fig.~\ref{fig-kagome-gaps}c), which remains stable over a wide range of parameter values.

Because quantum metric-dependent quantities enter at a higher order than Berry curvature fluctuations in the band geometry analysis, they should have a subdominant effect on the gap. In Fig.~\ref{fig-kagome-shells} we control for the effects of large curvature fluctuations by restricting attention to parameter values giving near-minimal values of $\sigma_B$. Including all such parameters yields a one-way relationship for the Laughlin and Moore-Read states (Figs.~\ref{fig-kagome-shells}a,~c), in the sense that large gaps are obtained only at low values of $\langle T\rangle$, but small gaps can be obtained at any value of $\langle T\rangle$. Further detail is evident if we take parameter values chosen to give the same value of $\sigma_{B}$ (see the Methods for a description of the sampling procedure used.) Results for sets of points chosen to have four different values of $\sigma_{B}$ are shown in Figs.~\ref{fig-kagome-shells}b,~d for the Laughlin and Moore-Read states. Removing the variation in $\sigma_B$ reveals a full-fledged negative correlation between $\langle T\rangle$ and the gaps: the trend is approximately linear for models having curvature fluctuations near the minimum and becomes less so as curvature fluctuations are allowed to increase.

\phantom{}\vfill

\subsection*{Ruby lattice model}
\label{sec-ruby}

Hu, Kargarian and Fiete \cite{Hu:2011br} described a Chern insulator model on the ruby lattice (Fig.~\ref{fig-ruby}a). In the limit of total spin polarization, they showed that hopping parameters could be chosen such that the lowest band had $c_1=1$ and a bandgap to bandwidth ratio of $\sim70$. The Hamiltonian for this model is
\begin{align}
\label{eq_ruby_ham}
H_\text{R} (\bfk) &=- t \sum_{j=1}^3 \left(e^{-i \bfk \cdot \bfa_{j}} h_{2j+1, 2j-1}(\bfk)+ h_{2j+2,2j}(\bfk)\right) \nonumber \\
{}&\quad -  t_1 \sum_{j=1}^3 \left( h_{2j,2j-1}(\bfk) + e^{i \bfk \cdot \bfa_{j+1}}h_{2j+1,2j}(\bfk) \right) \nonumber \\
{}&\quad - t_4 \sum_{j=1}^3 \left( h_{2j+3,2j}(\bfk) + e^{-i \bfk \cdot \bfa_{j+2}} h_{2j,2j-3}(\bfk) \right) \nonumber \\
{}&\quad + \text{h.c. } 
\end{align}
where $h_{b,c}(\bfk)$ is defined in eq.~(\ref{eq_hopping_def}). Here $\bfa_3=-\bfa_1-\bfa_2$ and the index on $\bfa$ is interpreted cyclically mod 3, but the indices on $h_{b,c}(\bfk)$ are interpreted cyclically mod 6. The relative offsets $\bfd_b$ are as depicted by the numbered sites in Fig.~\ref{fig-ruby}a. We considered the $c_1=1$ phase containing the flat-band point found in ref.~\onlinecite{Hu:2011br}, with $t=1.0+1.2i$, $t_1=-1.2+2.6i$, and $t_4=-1.2$. Momentum dependence of the Berry curvature and quantum metric is shown in Fig.~\ref{fig-ruby}b for parameters which minimize $\sigma_{B}$ and maximize the gaps for the Laughlin and Moore-Read states; these values are listed in Supplementary Table~4. Remarkably, the complexity of this Hamiltonian works in our favor: one can find parameter values which greatly reduce the fluctuations in band geometry relative to the kagom\'e lattice model, which means that the ruby lattice model may be tuned to produce a much closer approximation to lowest Landau level physics.

As a consequence, trends identified in the kagom\'e lattice model are manifest here with a much higher degree of correlation. Figs.~\ref{fig-ruby}c,~d show that the gaps of the Laughlin and Moore-Read states are strongly correlated with band geometry as measured by both $\sigma_{B}$ and $\langle T\rangle$. In both cases the gap can be seen to decrease with increasing $\langle T\rangle$, even for the same values of $\sigma_{B}$; in particular, the Moore-Read state is only stable in the lower right half of the plot area. Restricting our attention to parameters yielding small fluctuations in Berry curvature, in Fig.~\ref{fig-ruby-shells} we display data analogous to that presented for the kagom\'e lattice model in Fig.~\ref{fig-kagome-shells}. Again, if we only impose an upper bound on $\sigma_{B}$ we see that it's not possible to have a large gap for large values of $\langle T\rangle$ (Figs.~\ref{fig-ruby-shells}a,~c), while upon restricting to specific values of $\sigma_{B}$ (Figs.~\ref{fig-ruby-shells}b,~d) we see that the negative correlation between $\langle T\rangle$ and the gaps is even tighter, becoming nearly linear for low $\sigma_{B}$ and gaining more scatter as $\sigma_{B}$ is increased.

\begin{widetext}

\begin{figure}[H]
\centering
\includegraphics[width=\TwoColumnWidth]{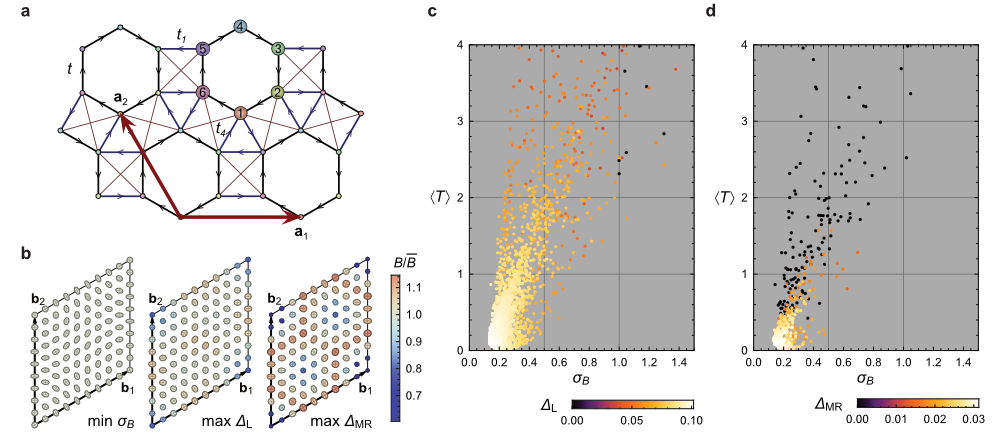}%
\caption{ 
\setstretch{\CaptionLineSpacing}
\textbf{Definition of couplings, band geometry and gaps for the ruby lattice model.} 
(\textbf{a}) Lattice and chirality conventions for hopping terms in the ruby lattice model Hamiltonian. All nearest-neighbor (arrowed) bonds are the same length. Basis vectors $\bfa_{1}, \bfa_{2}$ are shown in red; basis elements are shown by differently colored/numbered sites. Hopping elements are shown with black ($t$), blue ($t_{1}$) and brown ($t_{4}$) edges; arrowheads indicate the chirality convention for complex hoppings. (\textbf{b}) Band geometry over the reciprocal lattice unit cell spanned by $\bfb_{1}, \bfb_{2}$, for parameter values minimizing $\sigma_B$ and maximizing gaps, respectively. Axes of ellipses are proportional to the eigenvectors of the quantum metric $g^\alpha (\bfk)$. Ellipse color is given by the relative deviation of Berry curvature $B_\alpha(\bfk)$ from its Brillouin zone-averaged value. (\textbf{c}) Gap $\Delta$ as a function of the average Berry curvature fluctuation $\sigma_{B}$ and averaged trace condition $\langle T \rangle$ for the Laughlin state of $N=8$ bosons at $\nu = 1/2$.  (\textbf{d}) Gap as a function of $(\sigma_{B}, \langle T \rangle )$ for the Moore-Read state of $N=10$ bosons at $\nu = 1$.
}
\label{fig-ruby}
\end{figure}

\begin{figure}[H]
\centering
\includegraphics[width=\TwoColumnWidth]{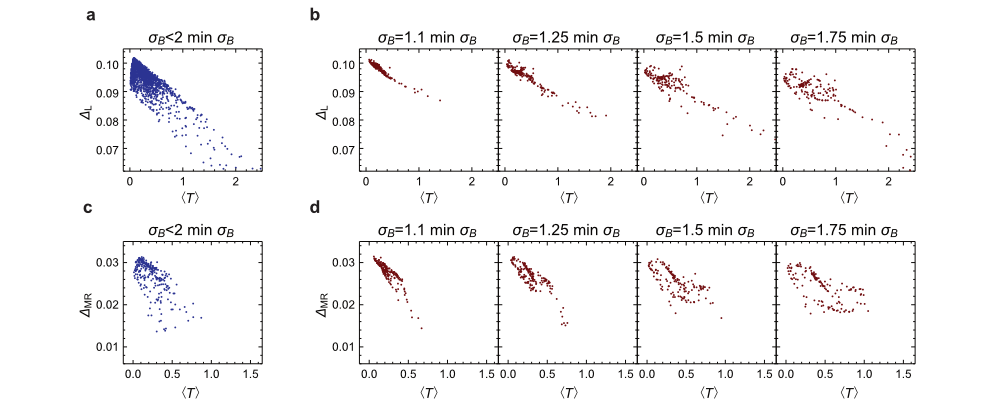}%
\caption{
\setstretch{\CaptionLineSpacing}
\textbf{Gap vs. trace condition for the ruby lattice model subject to constraints on $\sigma_{B}$.}  (\textbf{a}) Gaps for the bosonic Laughlin state of $N=8$ bosons at $\nu = 1/2$, as a function of the Brillouin zone average of the trace condition $\langle T \rangle$. We only plot gaps for parameter values which have Berry curvature fluctuations $\sigma_{B}$ less than twice its minimum value. (\textbf{b}) Gap of the bosonic Laughlin state vs. $\langle T \rangle$ for parameter values randomly chosen on isosurfaces of constant $\sigma_B$ in the space of couplings. (\textbf{c}),~(\textbf{d}) The same, for the bosonic Moore-Read state of $N=10$ bosons at $\nu = 1$. Note that the same sets of model parameters are used in each column.
}
\label{fig-ruby-shells}
\end{figure}

\end{widetext}

\subsection*{Significance of correlations}

In this section we describe two approaches to quantifying the degree of correlation between the band geometry and the many-body gaps described above. In particular, the fact that FQH-type states are destabilized by fluctuations in Berry curvature  is readily apparent and was anticipated in the first work on FCIs; in the present work we are interested in possible additional dependence on conditions derived from the quantum metric, so we seek to measure correlation between the gap and the trace condition $\langle T\rangle$. This is not fully straightforward, due to correlation of $\langle T\rangle$ and $\sigma_B$ with each other, as evidenced by the fact that parameter values are not uniformly distributed in Figs.~\ref{fig-kagome-gaps}c and \ref{fig-ruby}c. 

One approach is to compare data for parameters yielding the same value of $\sigma_B$, some of which is shown in Figs.~\ref{fig-kagome-shells}b,~d and \ref{fig-ruby-shells}b,~d (see the Methods for the procedure used to sample from isosurfaces of constant $\sigma_B$). We do not have quantitative predictions for the functional dependence of the  gap on any band-geometric quantity, so to avoid introducing assumptions we use Spearman's $\rho$ as a nonparametric measure of correlation. This is defined as the linear (Pearson) correlation coefficient between the rankings of the data points when rank-ordered by $\Delta$ and by $\langle T\rangle$, and takes values ranging from $+1$ for any monotonically increasing function to $-1$ for any monotonically decreasing function.

\begin{figure}[bh]
\centering
\includegraphics[width=\OneColumnWidth]{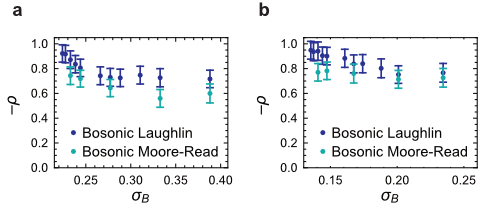}%
\caption{
\setstretch{\CaptionLineSpacing}
\textbf{Correlation between gaps and the trace condition when the magnitude of Berry curvature fluctuations is held constant.} Each data point shows (the negative of) Spearman's $\rho$ for the correlation between gap and averaged trace condition $\langle T \rangle$ on separate samples of $n =200$ points sampled from isosurfaces of constant root-mean-square Berry curvature fluctuation $\sigma_B$. A value of $\rho = -1$ corresponds to perfectly monotonic anti-correlation. We show results for Laughlin (dark blue) and Moore-Read (light blue) states in (\textbf{a}) the kagom\'e lattice model and (\textbf{b}) the ruby lattice model. Error bars show the bootstrapped standard error.
}
\label{fig-correls} 
\end{figure}

In Fig.~\ref{fig-correls} we plot the results of this test for the Laughlin and Moore-Read data in the kagom\'e and ruby lattice models. This score was found to be within $\sim 10\%$ of the linear correlation between $\langle T\rangle$ and $\Delta$ for each isosurface, which indicates the robustness of our conclusion and implies that all values of $\sigma_{B}$ considered lie in a weak-fluctuation regime. The general trend evident in Fig.~\ref{fig-correls} is that $\Delta$ is highly negatively correlated with $\langle T\rangle$ when curvature fluctuations are constrained to be near their minimum value. As curvature fluctuations are allowed to increase, this relationship becomes less exact, but converges to an asymptotic value well above 0 for both the Laughlin and Moore-Read states. This confirms the qualitative picture evident in Figs.~\ref{fig-kagome-shells}b,~d and \ref{fig-ruby-shells}b,~d.

Alternatively, we can analyze the larger set of data having unrestricted values of $\sigma_{B}$, at the cost of assuming linear relationships between all variables; the similarity between the results for Spearman's and Pearson's $\rho$ mentioned above suggest that this is justified. One can then compute the partial correlation, denoted here by $\rho_{B}$, as the degree of linear correlation between the residuals of $\Delta$ and $\langle T\rangle$, after first subtracting the best-fit linear dependence of each on $\sigma_B$; similarly, this score ranges from $+1$ to $-1$. For the augmented Haldane, kagom\'e and ruby lattice models, respectively, we find $\rho_{B}=-0.91,-0.42$ and $-0.62$ for the Laughlin state and $\rho_{B}=-0.77,-0.44$ and $-0.46$ for the Moore-Read state. Sample sizes were, respectively, $n=1900, 2100$ and $1800$. The fact that these values are lower than those obtained by the isosurface method describes the non-negligible correlation between $\sigma_B$ and $\langle T\rangle$. 

All correlation scores quoted above and shown in Fig.~\ref{fig-correls} are statistically significant at the $1\%$ level (at most). This is the main result of this section: gaps for both states are independently sensitive to variations in the trace condition, beyond the correlations induced in the latter via curvature fluctuations. This confirms that band geometry plays a significant role in realistic FCI models.

\subsection*{Cross-model comparisons}

The band geometry hypothesis claims that the most stable FQH-like phases are obtained when the band geometry is tuned to be as close to that of a Landau level as possible. We have shown above that this holds for several different lattice models as their couplings are varied, but comparison of the models shows an apparent contradiction: the gaps reported above are smallest for the ruby lattice model, despite the fact that this model can be made to approximate Landau level physics more closely than the other models studied here.

The resolution of this apparent contradiction lies in the fact that the models considered have different numbers of tight-binding sites per unit cell. This factor enters into the interaction term of the Hamiltonian, and hence the gap: because we have flattened the dispersion of the kinetic term, the interaction strength is the only energy scale in the problem. In Supplementary Note~4 we give a scaling argument that the strength of a delta-function interaction in the continuum should be multiplied by a factor of $\calN$ when discretized to an on-site repulsion in a model with $\calN$ bands. This means that, when comparing the gaps of bosonic Laughlin states in different models, the appropriate quantity to compare is $\calN \Delta$. For the three-body delta-function interaction which stabilizes the bosonic Moore-Read state, the same considerations yield a scaling factor of $\calN^{2}$.

\begin{widetext}

\begin{figure*}[hb]
\includegraphics[width=\TwoColumnWidth]{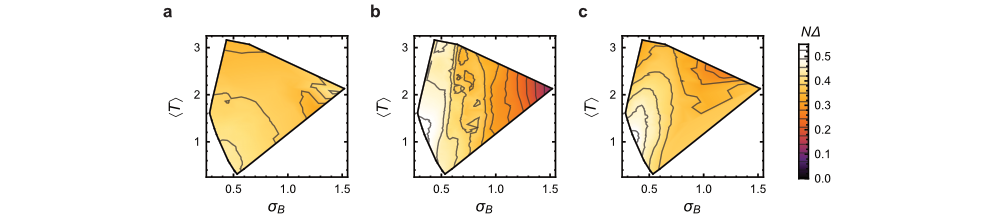}%
\caption{
\setstretch{\CaptionLineSpacing}
\textbf{Cross-model comparison of scaled gaps as a function of band geometry.} Panels show interpolated energy gaps of the bosonic Laughlin state, scaled by the number of bands of each model, for the (\textbf{a}) augmented Haldane, (\textbf{b}) kagom\'e lattice and (\textbf{c}) ruby lattice models described above. Scaled gaps are plotted as a function of band geometry as measured by Berry curvature fluctuations $\sigma_{B}$ and the average value of the trace condition $\langle T \rangle$. Data are only plotted in the region of common overlap of the three models in $(\sigma_B, \langle T \rangle)$ space (polygonal outline). The color scale and contours used are identical for all three panels.
}
\label{fig-crossmodel} 
\end{figure*}

\end{widetext}

In Fig.~\ref{fig-crossmodel} we compare the gaps for the Laughlin state for the augmented Haldane, kagom\'e and ruby lattice models when scaled by this factor, as a function of the band-geometric parameters $(\sigma_B,\langle T\rangle)$. The domain of each plot is restricted to be the common overlap of the three models in $(\sigma_B,\langle T\rangle)$ space. Because of this restriction, we were unable to perform a meaningful analysis for the Moore-Read state, as it is unstable in most of this region (compare Figs.~\ref{fig-kagome-gaps}d, \ref{fig-ruby}d). Despite the fact that the scaling argument is exact only in the large-$\calN$ limit and the models we compare have $\calN=2,3$ and $6$, we find roughly similar behavior across all three models, both in terms of the magnitude of the scaled gap and of its dependence on $\sigma_B$ and $\langle T\rangle$.

\section*{Discussion}

In this work we have presented numerical results which systematically map out the robustness of topologically ordered FQH-like phases in FCI systems with realistic, short-ranged Hamiltonians and band geometry which is less than perfectly uniform. We presented quantitative evidence that the band-geometric quantities identified in ref.~\onlinecite{Roy:2012vo} remain strongly correlated with the size of the gap even when its conditions on band geometry are not met exactly. This leads us to propose a geometric stability hypothesis for FQH-like phenomena in FCIs: in spite of the fact that the GMP algebra is not perfectly reproduced, we conjecture that an approximate version of the single-mode approximation correctly describes the low-energy physics of these FCI models. As a practical corollary, we predict that single-particle Hamiltonians with more uniform band geometry --- specifically, as measured by the hierarchy of three criteria --- will produce more stable FQH-like states.

The validity of the single-mode approximation in FCIs has been investigated by a number of other authors using approaches complementary to that taken here\cite{Parameswaran:2012cu,Goerbig:2012cz,Roy:2012vo,Dobardzic:2013gc}. Within the context of their Hamiltonian approach to the FQHE, Murthy and Shankar showed that composite fermion degrees of freedom could be chosen which reproduce an exact version of the GMP algebra\cite{Murthy:2011vi,Murthy:2012ta}. More directly, in ref.~\onlinecite{Repellin:2014fn}, the lowest-lying neutral excitation of the kagom\'e and ruby lattice models was found to be well described by the magnetoroton mode of the corresponding FQH state on a torus, using a phenomenological mapping between the FQH and FCI Hilbert spaces described in ref.~\onlinecite{Bernevig:2012ka}. A related mapping was originally proposed by Qi\cite{Qi:2011jo}, but the image of FQH pseudopotential interactions under this mapping is not well localized and strongly anisotropic\cite{Qi:2011jo,Wu:2012eq,Wu:2013ii,Lee:2013bz,Lee:2013wd}, making the relationship to physical FCI Hamiltonians unclear.

Comparing the results from the kagom\'e and ruby lattice models, it appears easier to engineer uniform geometry in more complicated Hamiltonians, both in the sense of having more tunable couplings and in the sense of having more bands. The latter property is expected to hold on general grounds, as noted in refs.~\onlinecite{Cooper:2012bt,Kovrizhin:2013ib}. Increasing the size of the unit cell reduces the effectiveness of a fixed-strength repulsion, however, so an optimal choice would balance these two factors. This has immediate relevance to experimental design: laboratory Hamiltonians are necessarily more complicated than those in idealized theoretical models (e.g., the proposal in ref.~\onlinecite{Yao:2013eg} involves an eight-dimensional parameterization of the applied electric field used to obtain a synthetic gauge potential.) Performing many-body simulations on a representative set of parameters in such a large space is prohibitively time-consuming; the geometric stability hypothesis can be used to reduce this to a manageable subspace. In addition, band geometry may, by definition, be tuned independently of energetic considerations such as the bandwidth. 

A pressing direction for future work is to further develop the band geometry hypothesis by investigating its validity in less straightforward scenarios: stable FCI states where the Berry curvature is not particularly uniform have also been proposed; furthermore, the stability of the state may also depend on the filling fraction and the particular state sought to be stabilized\cite{Kapit:2010ky,Liu:2013fo}. Among other aspects which would be interesting to clarify are the role of bosonic versus fermionic particle statistics, NN- and longer-ranged inter-particle interactions, anisotropic interactions induced by the lattice structure, etc. In particular, the distribution of geometric quantities differs from band to band; this can be selected by fully filling a number of bands in a fermionic system, which could lead to new phenomena. One could also consider the wide range of more elaborate FCI models in the literature, possessing, e.g., Chern numbers $|c_1|>1$, non-Abelian Berry curvature arising from multiple degenerate bands, etc. We direct the reader to the recent reviews\cite{Parameswaran:2013uf,Bergholtz:2013ue} for a more extensive discussion and bibliography.

\section*{Methods}
\label{sec-methods}

\subsection*{Parameter space sampling}

We study the dependence of the many-body gap on band-geometric quantities in the Haldane model \cite{Haldane:1988gh} and models proposed for the kagom\'e \cite{Tang:2011by} and ruby \cite{Hu:2011br} lattices. The parameter space for the Haldane model Hamiltonian is small enough that the vector of parameters $\bft_\Delta$ which maximize the gap $\Delta$ may be found from exhaustive sampling. The parameter spaces for the other two models are higher-dimensional and non-compact, so this strategy will not work.

According to the geometric stability hypothesis, the Berry curvature fluctuations $\sigma_B$ should be inversely correlated with the gap, meaning that it may be employed as a proxy for the latter which requires far less effort to compute. In Supplementary Note~5 we derive an expression for the derivative of $\sigma_B$ with respect to any parameter appearing in the single-particle Hamiltonian, allowing steepest-descent methods to be employed. Of course, we do not expect the vector of parameters $\bft_0$ minimizing $\sigma_B$ to precisely coincide with those maximizing the gap, but if the single mode approximation is applicable, $\bft_0$ will be a viable initial guess for $\bft_\Delta$, to be refined as described below. Indeed, a state with a robust topological gap in the presence of large curvature fluctuations would be of immediate interest as an FCI phase not describable in terms of an FQH universality class. 

\begin{figure}[ht]
\includegraphics[width=\OneColumnWidth]{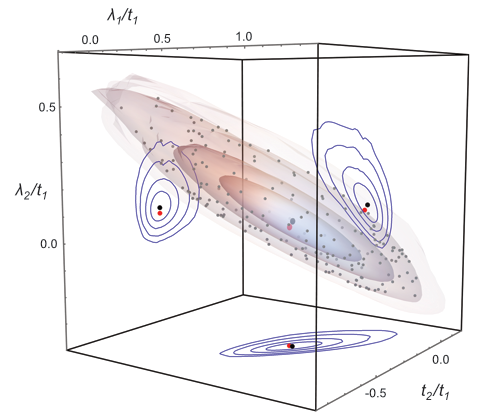}%
\caption{
\setstretch{\CaptionLineSpacing}
\textbf{Isosurfaces of constant Berry curvature fluctuations in the parameter space of the kagom\'e lattice model.}
Parameter values which minimize the root-mean-square Berry curvature fluctuation $\sigma_{B}$ of the kagom\'e lattice model are marked by the large, central black dot. The large red dot identifies parameter values which yielded the maximum gap for the Laughlin state of $N=8$ bosons at $\nu = 1/2$. Concentric shaded surfaces are isosurfaces upon which $\sigma_{B}$ takes a constant value equal to $1.05, 1.1, 1.25$ and $1.5$ times its minimum value (respectively). Blue contours on the walls of the box are sections through these surfaces at the  minimum $\sigma_B$ point. Small grey points identify random samples taken from the $\sigma_B = 1.25 \min \sigma_B$ isosurface, to illustrate uniformity.
}
\label{fig-kag-samples} 
\end{figure}

In the neighborhood of $\bft_0$, a surface of constant $\sigma_B$ in parameter space will be approximated by an ellipsoid given by the Hessian of $\sigma_B(\bft)$ at $\bft_0$, which may be calculated by the method in Supplementary Note~5. We sample points uniformly from the surface of the ellipsoid defined by the Hessian by the well-known method of projecting vectors of parameters $\bft$ sampled from the corresponding multinormal distribution. We then perturb this ellipsoid by shifting the vectors $\bft$ radially along the rays connecting each with $\bft_0$ until we find parameters $\bft'$ such that $\sigma_B(\bft')$ is equal to the target value. These isosurfaces of different parameters with the same value of $\sigma_B$ permit us to study the sub-leading effects of the quantum metric on the gap predicted by the geometric stability hypothesis: in Fig.~\ref{fig-kag-samples} we depict the isosurfaces found for the kagom\'e lattice model which were used to generate the data shown in Figs.~\ref{fig-kagome-shells}b,~d.

Finally, we locate the parameters giving the maximum gap by fitting a quadratic form
\begin{equation}
\Delta(\bft) = \Delta^\ast - \frac{1}{2} \sum_{i,j} \Sigma_{ij}^{-1} (t_i - t_i^\ast)  (t_j - t_j^\ast)
\end{equation}
to the gaps from the isosurface data and sampling new parameters from a Gaussian distribution centered on $\bft^\ast$ with covariance matrix $\Sigma$. We have verified that deviations of the actual $\bft_\Delta$ found from this data from the fitted value $\bft^\ast$ are negligible compared to the scales set by $\Sigma$, hence this search does not need to be iterated further.

\subsection*{Numerical exact diagonalization}

Unless explicitly identified otherwise, all data were obtained from exact diagonalization of the many-body Hamiltonian for $N=8$ bosons interacting with a two-body on-site repulsion at a filling fraction (ratio of the number of particles to the number of unit cells) of $\nu = 1/2$ on a periodic lattice of $4 \times 4$ unit cells (for the Laughlin-like FCI state) or $N=10$ bosons at a filling fraction of $\nu = 1$ on a lattice of $5 \times 2$ unit cells, interacting with a three-body on-site repulsion only (for the Moore-Read-like FCI state). In the main text we also presented additional data obtained for the fermionic Laughlin-like state in the Haldane model with $N=8$ particles at $\nu = 1/3$ on a $6 \times 4$ lattice with two-body nearest-neighbor repulsion. The fact that the bosonic Moore-Read state is realized in a filled Landau level means that, relative to the Laughlin state, we are able to simulate more particles with a many-body Hilbert space of roughly the same size. Flattening the spectrum of the single-particle Hamiltonian removes the only other energy scale from the problem, so the strength of the repulsive interaction sets the units of the many-body gap $\Delta$. 

Because we do simulations for thousands of parameter values, a detailed finite-size scaling analysis of each is beyond the scope of this paper. We have, however, made spot checks by doing simulations of $N=6$ to $N=12$ particles at select parameter values. Representative data is shown in Supplementary Figure~6. From this analysis, we have concluded that the system sizes described above strike an acceptable balance between accurately approximating the thermodynamic ($N \to \infty$) limit and making the volume of simulations manageable on the computational resources available to us.

We employ several criteria to identify topological order in the results of numerical simulations, employing information from both the many-body energy spectrum and the many-body wavefunctions themselves. We first require that the spectrum has the correct number of degenerate ground states in the proper momentum sectors (two states at $(k_x, k_y) = (0,0)$ for bosonic Laughlin-type order, and three states at $(k_x, k_y) = (0,0), (0,0)$ and $(0,1)$ for bosonic Moore-Read-type order). In a finite-size simulation these ground states will only be approximately degenerate; we require that the gap to the lowest-energy excited state (out of all momentum sectors; i.e. the indirect gap) be at least as large as the spread in ground state energies. 

We ascertain whether a ground state wavefunction has the topological order of the Laughlin or Moore-Read state through properties of its entanglement spectrum \cite{Li2008}: we first verify that the spectrum of the reduced density matrix obtained by tracing out four bosons is gapped. There is no current quantitative theoretical interpretation of the magnitude of the entanglement gap; instead, the discriminative power of this criterion comes from requiring that the number of eigenvalues below the gap in each momentum sector obeys counting rules dictated by the topological order of continuum FQH state \cite{Regnault:2011bu,Bernevig:2012ka}. For the lattice geometry used in our studies, Laughlin-type order in a system of $N=8$ bosons at $\nu= 1/2$ is identified by the $(1,2)$ counting rule (in the terminology of the previous references), which requires $48$ states below the gap at $(k_x, k_y) = (0,0)$, $44$ states below the gap in sectors $(k_x, k_y) = (0,2), (2,0)$, and $(2,2)$, and $40$ states below the gap in all other momentum sectors. Similarly, the counting rule for Moore-Read-type order in $N=10$ bosons at $\nu= 1$ requires $76$ states below the gap at $(k_x, k_y) = (0,0)$ and $(0,1)$, and $75$ states below the gap in all other momentum sectors. Simulations that fail any of the above tests are assigned an energy gap of zero.





\vfill 

\begin{acknowledgments}
The authors thank Abhishek Roy for preliminary code written in the initial stage of the project and useful discussions and Emil Bergholtz for useful comments on a draft of this manuscript. R.~R. acknowledges support from the Sloan Foundation. G.~M. acknowledges support from the Leverhulme Trust under grant no.~ECF-2011-565, from the Newton Trust of the University of Cambridge, and from the Royal Society under grant UF120157. This work used computational and storage services associated with the Hoffman2 Shared Cluster provided by UCLA Institute for Digital Research and Education's Research Technology Group. Part of our numerical work was performed using the Darwin Supercomputer of the University of Cambridge High Performance Computing Service funded by Strategic Research Infrastructure Funding from the Higher Education Funding Council for England and funding from the Science and Technology Facilities Council.
\end{acknowledgments}

\onecolumngrid
\appendix

\renewcommand{\appendixname}{Supplementary Note}
\renewcommand{\figurename}{Supplementary Figure}
\renewcommand{\tablename}{Supplementary Table}

\renewcommand{\thesubsection}{\arabic{subsection}}
\renewcommand{\theequation}{\arabic{equation}}
\renewcommand{\thefigure}{\arabic{figure}}
\renewcommand{\thetable}{\arabic{table}}

\setcounter{figure}{0} 


\clearpage
\section*{Supplementary Figures}

\begin{figure}[hbt]
\centering
\includegraphics[width=\OnePointFiveColumnWidth]{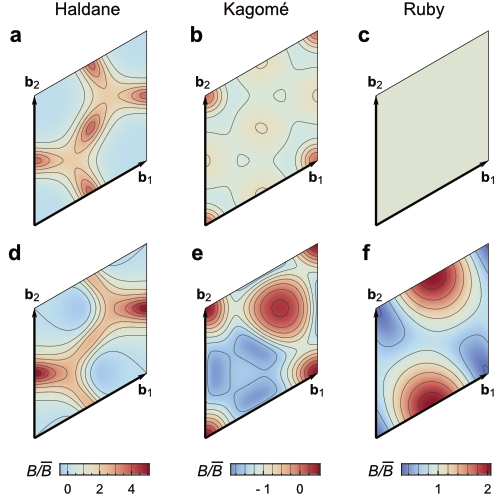}%
\caption{
\setstretch{1.25}
\textbf{Effect of Hamiltonian transformation on momentum space distribution of Berry curvature.} We plot the deviations from uniform Berry curvature (neutral grey color) over a unit cell of the reciprocal lattice spanned by $\bfb_{1}, \bfb_{2}$, for models treated in the main text. Plots in each column use the same color scale and set of contours, as indicated. (\textbf{a}) -- (\textbf{c}) Berry curvature distribution for the Haldane, kagom\'e lattice and ruby lattice models, respectively, at parameter values which minimize the root-mean-square curvature fluctuation $\sigma_B$ over the unit cell. Curvature fluctuations of the ruby lattice model at its $\min \sigma_{B}$ point (\textbf{c}) are too small to be visible on the common scale. (\textbf{d}) -- (\textbf{f}) Berry curvature distribution for the transformed versions of these models used in Supplementary Ref.~\onlinecite{Wu:2012do}, evaluated the parameter values minimizing the values of $\sigma_B$ as computed with the transformed Hamiltonians. }
\label{fig-zoo} 
\end{figure}

\begin{figure}[ht]
\centering
\includegraphics[width=\OnePointFiveColumnWidth]{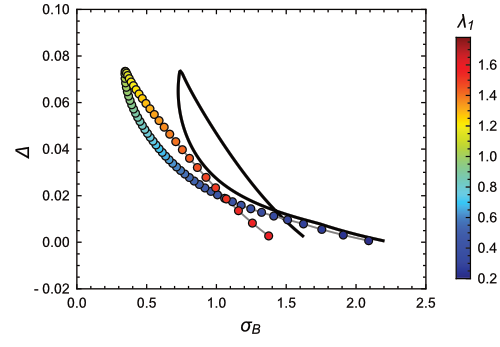}%
\caption{
\setstretch{1.25}
\textbf{Effect of Hamiltonian transformation on Berry curvature-gap correlations.}
Representative plot of the correlation between the many-body gap $\Delta$ and root-mean-square Berry curvature fluctuation $\sigma_B$. Data shown are for the Laughlin state of $N=8$ fermions in the kagom\'e lattice model with nearest-neighbor repulsion on a $6 \times 4$ lattice. The single-particle parameters used are $t_2 = \lambda_2 = 0$ and $t_1 = 1$, with the color scale corresponding to $\lambda_1$. This model and parameters are chosen to match those used in Supplementary Ref.~\onlinecite{Wu:2012do} (compare Fig.~26 of that reference); the thick black line corresponds to values of $\sigma_{B}$ computed using the transformed Hamiltonian of that reference, while colored circles denote values of $\sigma_{B}$ computed using the Hamiltonian in the main text. The two sets of data have the same gaps; only the computed values of $\sigma_{B}$ differ.}
\label{fig-zoo26} 
\end{figure}

\begin{figure}[ht]
\centering
\includegraphics[width=\OnePointFiveColumnWidth]{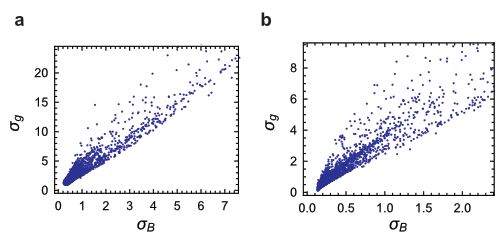}%
\caption{
\setstretch{1.25}
\textbf{Correlation between fluctuations of Berry curvature and of the quantum metric.} 
The root-mean-square (RMS) fluctuation of the quantum metric over the Brillouin zone ($\sigma_{g}$) is plotted against the RMS fluctuation of the Berry curvature $\sigma_{B}$, for all parameter values used in the text for the (\textbf{a}) the kagom\'e lattice model and (\textbf{b}) the ruby lattice model. Despite sampling a large volume of the parameter space of both models, the fluctuations in both quantities show a large degree of linear correlation.
}
\label{fig-sigmas} 
\end{figure}

\begin{figure}[ht]
\centering
\includegraphics[width=\TwoColumnWidth]{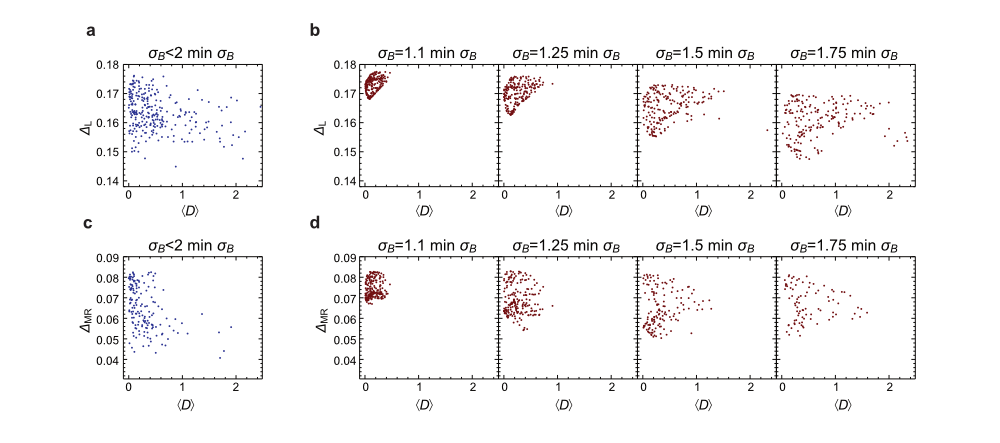}%
\caption{
\setstretch{1.25}
\textbf{Gap vs. determinant condition for the kagom\'e lattice model subject to constraints on $\sigma_{B}$.} 
(\textbf{a}) Gaps for the bosonic Laughlin state of $N=8$ bosons at $\nu = 1/2$, as a function of the Brillouin zone average of the determinant condition $\langle D \rangle$. We only plot gaps for parameter values which have Berry curvature fluctuations $\sigma_{B}$ less than twice its minimum value. (\textbf{b}) Gap of the bosonic Laughlin state vs. $\langle D \rangle$ for coupling values randomly chosen on isosurfaces of constant $\sigma_B$ in the space of couplings. The parameter space sampling procedure used to obtain these sets of points is described in the main text (see Methods). (\textbf{c}),~(\textbf{d}) The same, for the bosonic Moore-Read state of $N=10$ bosons at $\nu = 1$. Note that the same sets of couplings are used in each column. 
}
\label{fig-k-detineq}
\end{figure}

\begin{figure}[ht]
\centering
\includegraphics[width=\TwoColumnWidth]{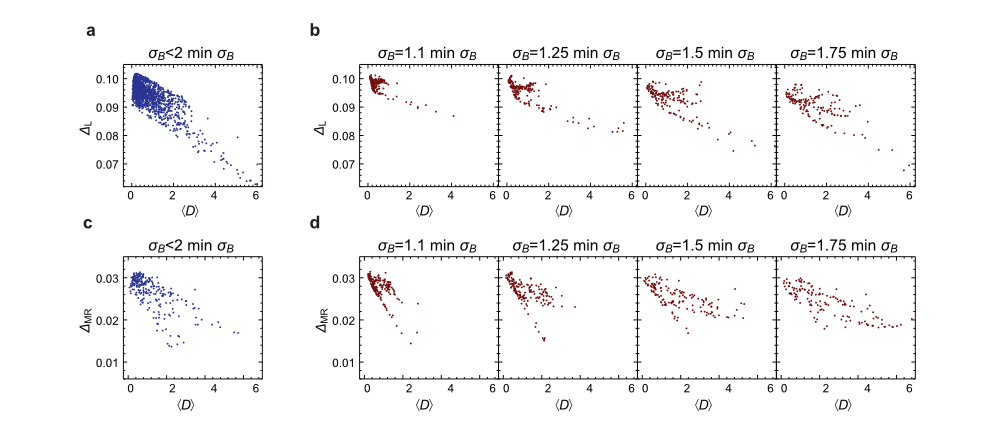}%
\caption{
\setstretch{1.25}
\textbf{Gap vs. determinant condition for the ruby lattice model subject to constraints on $\sigma_{B}$.}  
(\textbf{a}) Gaps for the bosonic Laughlin state of $N=8$ bosons at $\nu = 1/2$, as a function of the Brillouin zone average of the determinant condition $\langle D \rangle$. We only plot gaps for parameter values which have Berry curvature fluctuations $\sigma_{B}$ less than twice its minimum value. (\textbf{b}) Gap of the bosonic Laughlin state vs. $\langle D \rangle$ for coupling values randomly chosen on isosurfaces of constant $\sigma_B$ in the space of couplings. The parameter space sampling procedure used to obtain these sets of points is described in the main text (see Methods). (\textbf{c}),~(\textbf{d}) The same, for the bosonic Moore-Read state of $N=10$ bosons at $\nu = 1$. Note that the same sets of couplings are used in each column.
}
\label{fig-r-detineq}
\end{figure}

\begin{figure}[ht]
\centering
\includegraphics[width=\OnePointFiveColumnWidth]{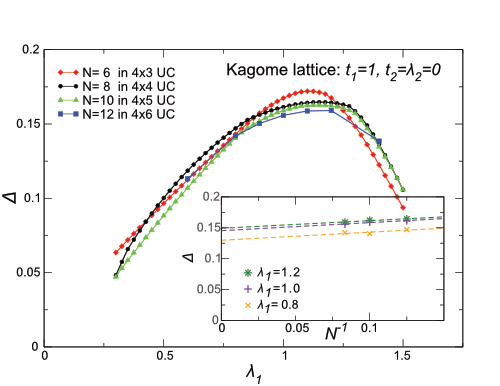}%
\caption{
\setstretch{1.25}
\textbf{Finite-size scaling of the many-body gap in the kagom\'e lattice model with nearest-neighbor hoppings.} 
The many-body gap data presented in the main text were obtained for systems of $N=8$ particles, which represents a compromise between the need to accurately estimate the gap in thermodynamic limit and the need to exhaustively sample the parameter space of single-particle Hamiltonians using the computational resources available to us. Here we plot the many-body gap $\Delta$ for the bosonic Laughlin state in the  kagom\'e lattice model with nearest-neighbor hopping $t_1 = 1$, variable  $\lambda_1$, and next-nearest-neighbor hopping $t_2=\lambda_2 = 0$. Red, black, green and blue points denote data for systems with $N=6$, $8$, $10$ and $12$ particles, respectively: the behavior changes significantly between system sizes $N=6$ and $N=8$, but the gap data for $N=10$ and $N=12$ closely mirrors that of $N=8$, especially near the maximum of the gap. Inset: finite size scaling of the gap as a function of $1/N$ for selected values of $\lambda_1$. Near the maximal gap, the thermodynamic extrapolation ($N \to \infty$) for the gap differs from the finite size value at $N=8$ by about $10$\%.}
\label{fig-scaling}
\end{figure}

\clearpage
\section*{Supplementary Tables}

\begin{table}[H]
\caption{\label{tab-haldane} \textbf{Parameter values, band-geometric quantities and gaps for the Haldane model.}} \begin{ruledtabular}
\begin{tabular}{cxxx}
{} & \multicolumn{1}{c}{$\min \sigma_B$} &  \multicolumn{1}{c}{$\max \Delta_{\text{L}}, \text{ Fermions}$} &  \multicolumn{1}{c}{$\max \Delta_{\text{L}}, \text{ Bosons}$}  \\
\hline
$\phi$ & 0.225 & 0.131 & 0.354 \\
$M$ & 0.0 & 0.0 & 0.0 \\
$\sigma_B$ & 0.99326 & 1.24755 & 1.12760 \\
$\sigma_g$ & 2.85734 & 3.97585 & 3.25586 \\
$\langle D \rangle$  & 0 & 0 & 0 \\
$\langle T \rangle$ & 0.46705 & 0.84597 & 0.99358 \\
$\Delta$ &  & 0.02554 & 0.17243 
\end{tabular}
\end{ruledtabular}

As in the main text, we take the hopping amplitudes to be $t_1 = t_2 = 1$, with the remaining couplings $\phi$, $M$ as free variables. Columns list information for the following values of $(\phi,M)$ of interest: those which minimize Berry curvature fluctuations $\sigma_B$ (obtained using the steepest-descent procedure described in Supplementary Note \ref{sec-gradients}) and those maximizing the many-body gap $\Delta$ for the Laughlin state for $N=8$ fermions at $\nu = 1/3$ and $N=8$ bosons at $\nu = 1/2$. The latter two parameter values were found using our parameter space sampling procedure (see Methods). For these parameters, we compute the root-mean-square average over the Brillouin zone of fluctuations in the Berry curvature ($\sigma_{B}$) and quantum metric ($\sigma_{g}$), as well as the Brillouin zone averages of the determinant ($\langle D \rangle$) and trace ($\langle T \rangle$) conditions.
\end{table}

\begin{table}[H]
\caption{\label{tab-haldane2} \textbf{Parameter values, band-geometric quantities and gaps for the augmented Haldane model with third-nearest neighbor hopping.}}
\begin{ruledtabular}
\begin{tabular}{cxxx}
{} & \multicolumn{1}{c}{$\min \sigma_B$} & \multicolumn{1}{c}{$\max \Delta_{\text{L}}$} & \multicolumn{1}{c}{$\max \Delta_{\text{MR}}$}  \\
\hline
$t_{2}$ & 1.0 & 0.55 & 0.75 \\
$t_{3}$ & -1.0 & -0.65 & -0.65 \\
$\sigma_B$ & 0.14699 & 0.43579 & 0.34909 \\
$\sigma_g$ & 3.00134 & 1.70829 & 2.29776 \\
$\langle D \rangle$  & 0 & 0 & 0 \\
$\langle T \rangle$ & 3.44154 & 0.75821 & 1.46273 \\
$\Delta$ & {} & 0.19153 & 0.12614
\end{tabular}
\end{ruledtabular}

As in the main text, we fix  $t_1=1$, $\phi = \pi/2$, and $M=0$, allowing $t_{2}$, $t_{3}$ to vary. Columns list information for the following values of $(t_{2},t_{3})$ of interest: those which minimize Berry curvature fluctuations $\sigma_B$ (obtained using the steepest-descent procedure described in Supplementary Note \ref{sec-gradients}) and those maximizing the many-body gap $\Delta$ for the Laughlin state of $N=8$ bosons at $\nu = 1/2$ and for the Moore-Read state of of $N=10$ bosons at $\nu = 1$. The latter two parameter values were found using our parameter space sampling procedure (see Methods). For these parameters, we compute the root-mean-square average over the Brillouin zone of fluctuations in the Berry curvature ($\sigma_{B}$) and quantum metric ($\sigma_{g}$), as well as the Brillouin zone averages of the determinant ($\langle D \rangle$) and trace ($\langle T \rangle$) conditions.
\end{table}

\begin{table}[H]
\caption{\label{tab-kagome} \textbf{Parameter values, band-geometric quantities and gaps for the kagom\'e lattice model.}}\begin{ruledtabular}
\begin{tabular}{cxxx}
{} & \multicolumn{1}{c}{$\min \sigma_B$} & \multicolumn{1}{c}{$\max \Delta_{\text{L}}$} & \multicolumn{1}{c}{$\max \Delta_{\text{MR}}$}  \\
\hline
$\lambda_1$ &  0.745 & 0.725 & 0.934 \\
$t_2$ & -0.361 & -0.361 & -0.168 \\
$\lambda_2$ & 0.078 & 0.055 & -0.129 \\
$\sigma_B$ & 0.22178 & 0.23231 & 0.34967 \\
$\sigma_g$ & 1.31746 & 1.25972 & 1.25434 \\
$\langle D \rangle$  & 0.09608 & 0.09857 & 0.41665 \\
$\langle T \rangle$ & 0.79612 & 0.70181 & 0.42712 \\
$\Delta$ & {} & 0.17614 & 0.08246
\end{tabular}
\end{ruledtabular}

We fix an energy scale by setting $t_{1} = 1$ and letting the remaining model parameters ($\lambda_{1}$, $t_{2}$, $\lambda_{2}$) vary. Columns list information for the following parameter values of interest: those which minimize Berry curvature fluctuations $\sigma_B$ (obtained using the steepest-descent procedure described in Supplementary Note \ref{sec-gradients}) and those maximizing the many-body gap $\Delta$ for the Laughlin state of $N=8$ bosons at $\nu = 1/2$ and for the Moore-Read state of of $N=10$ bosons at $\nu = 1$. The latter two parameter values were found using our parameter space sampling procedure (see Methods). For these parameters, we compute the root-mean-square average over the Brillouin zone of fluctuations in the Berry curvature ($\sigma_{B}$) and quantum metric ($\sigma_{g}$), as well as the Brillouin zone averages of the determinant ($\langle D \rangle$) and trace ($\langle T \rangle$) conditions.

\end{table}

\begin{table}[H]
\caption{\label{tab-ruby} \textbf{Parameter values, band-geometric quantities and gaps for the ruby lattice model.}} \begin{ruledtabular}
\begin{tabular}{cxxx}
{} & \multicolumn{1}{c}{$\min \sigma_B$} & \multicolumn{1}{c}{$\max \Delta_{\text{L}}$} & \multicolumn{1}{c}{$\max \Delta_{\text{MR}}$}  \\
\hline
$\Imag t $ & 1.857 & 1.886 & 1.672 \\
$\Real t_1 $ & 0.685 & -0.569 & -1.261 \\
$\Imag t_1 $ & 2.079 & 2.536 & 2.789 \\
$t_4 $ & -2.118 & -2.022 & -2.007 \\
$\sigma_B$ & 0.13398 & 0.14347 & 0.17980 \\
$\sigma_g$ & 0.47669 & 0.21574 & 0.38617 \\
$\langle D \rangle$ & 0.17740 & 0.25569 & 0.30456 \\
$\langle T \rangle$ & 0.17168 & 0.08262 & 0.11139 \\
$\Delta$ &{}& 0.10173 & 0.03132
\end{tabular}
\end{ruledtabular}

We fix an energy scale by setting $\Real t = 1$ and letting the remaining parameters ($\Imag t$, $t_{1}$, $t_{4}$) vary. Columns list information for the following parameter values of interest: those which minimize Berry curvature fluctuations $\sigma_B$ (obtained using the steepest-descent procedure described in Supplementary Note \ref{sec-gradients}) and those maximizing the many-body gap $\Delta$ for the Laughlin state of $N=8$ bosons at $\nu = 1/2$ and for the Moore-Read state of of $N=10$ bosons at $\nu = 1$. The latter two parameter values were found using our parameter space sampling procedure (see Methods). For these parameters, we compute the root-mean-square average over the Brillouin zone of fluctuations in the Berry curvature ($\sigma_{B}$) and quantum metric ($\sigma_{g}$), as well as the Brillouin zone averages of the determinant ($\langle D \rangle$) and trace ($\langle T \rangle$) conditions.
\end{table}

\clearpage
\section*{Supplementary Notes}

\renewcommand{\thesection}{\arabic{section}}

\section{Gauge freedom and geometric phases in band Hamiltonians}
\label{sec-nogauge}

Here we discuss issues arising in the definition of Berry curvature for Bloch bands which are absent from the general formalism of geometric phases. The issues identified here are not new\cite{Zak:1989fh,Zak:1989ia}; we discuss them here in order to demonstrate that the values the Berry curvature and quantum metric take at a specific crystal momentum $\bfk$ are unambiguously defined and, in principle, measurable.

Berry curvature arises in physical applications at the as the commutator of band-projected position operators; as examples, we cite the semiclassical approximation to the orbital magnetization\cite{Sundaram:1999ht} which contributes to anomalous thermoelectric transport, and the intrinsic contribution to the anomalous Hall conductance\cite{Karplus:1954bv}
\begin{equation}
\sigma_{xy}^{\text{int}} = \frac{e^2}{\hbar} \int \frac{d^dk}{(2 \pi)^d} \, f(E_{\bfk}) B(\bfk),
\end{equation}
Other applications are reviewed in Supplementary Ref.~\onlinecite{Xiao:2010kw}. We mention these observables here because they depend on the Berry curvature through forms \emph{other} than its Brillouin zone (BZ) average (the Chern number), meaning that the distribution of curvature within the BZ is an experimentally meaningful quantity. There is also a recent experimental proposal to measure the Berry curvature directly\cite{Duca:2014wh}.

Textbook discussions of Berry's phase are usually framed in the context of adiabatic evolution of a quantum state tracing out a closed cycle in some parameter manifold. In Chern insulator applications, the parameter manifold is the Brillouin zone, and the instantaneous eigenfunctions at a  parameter $\bfk$ are the spatially periodic part of the Bloch functions $u^\alpha_b(\bfk)$. The Berry connection is $\bfA_\alpha(\bfk) = -i \langle u^\alpha_\bfk | \nabla_{\bfk} | u^\alpha_\bfk \rangle$, which has curvature given by 
\begin{equation}
B_\alpha(\bfk) = -i \sum_{b=1}^{\calN} \left( \frac{\partial u_b^{\alpha \ast}}{\partial k_x}\frac{\partial u_b^{\alpha}}{\partial k_y} -  \frac{\partial u_b^{\alpha \ast}}{\partial k_y}\frac{\partial u_b^{\alpha}}{\partial k_x} \right).
\end{equation}
The first Chern number is defined as the surface integral of the Berry curvature
\begin{equation}
c_1 = \frac{1}{2\pi} \int_{BZ} d^2k \, B_\alpha (\bfk)
\end{equation}
and is topologically quantized to integer values due to the fact that the Brillouin zone is a compact manifold (a torus). 

Eigenstates of the band Hamiltonian
\begin{equation}
H_{bc}(\bfk) = \sum_{\alpha = 1}^{\calN} E_\alpha (\bfk) u^{\alpha \ast}_b(\bfk) u^\alpha_c(\bfk)
\end{equation}
are only defined up to an overall phase
\begin{equation}
\label{eq_bloch_gauge}
|u^\alpha_{\bfk} \rangle \to e^{i \phi_\alpha(\bfk)} |u^\alpha_{\bfk} \rangle, 
\end{equation}
where $\phi_\alpha(\bfk)$ is any smooth function satisfying $\phi_\alpha(\bfk + \bfG) = \phi_\alpha(\bfk)$. This is the gauge symmetry of the band Hamiltonian, and it is the only such symmetry in the absence of energy degeneracies (assumed for simplicity throughout this section). As with the $U(1)$ gauge symmetry of electromagnetism, gauge transformations of the form (\ref{eq_bloch_gauge}) alter the Berry connection but leave the Berry curvature (analogue of the magnetic field) and quantum metric unchanged, as can be seen from their explicitly gauge-invariant forms (\ref{eq_b_curvature}), (\ref{eq_q_metric}). 

The reader will note that the band Hamiltonians used in this article do \emph{not} have the periodicity of the reciprocal lattice; nor do their eigenfunctions $|u^{\alpha}_\bfk\rangle$, and so neither can immediately be viewed as functions on the BZ torus defined by identifying the points $\bfk$ and $\bfk + \bfG$ for any reciprocal lattice vector $\bfG$. We first explain that the formalism of band geometry is unchanged in this situation, and then argue that this choice of basis is the canonically correct one, in the sense of corresponding to the observable quantities mentioned above.

Because the full Hamiltonian is periodic in real space, Bloch's theorem implies 
\begin{equation}
u_{\bfk}^\alpha (\bfr=\bfd_b) \equiv \langle \mathbf{0} b | u_{\bfk}^\alpha \rangle = \sum_\bfG c_{\bfk-\bfG} e^{i\bfG \cdot \bfd_b}
\end{equation} 
is unchanged under $\bfd_b \to \bfd_b + \bfR$; in other words, there exists a unitary matrix $U_\bfG = e^{i \bfG \cdot \widehat{\bfr}}$ such that
\begin{equation}
u^\alpha_b (\bfk + \bfG) = \sum_{c=1}^\calN (U_\bfG)^{\phantom{\alpha}}_{bc} u^\alpha_c (\bfk),
\end{equation} 
for \emph{all} $\bfk, \alpha$. Because $U_\bfG$ is independent of $\bfk$, it drops out of the expressions for the Berry curvature and quantum metric, which are therefore periodic in $\bfk$.

One could also obtain manifestly periodic Bloch functions by performing momentum-dependent phase shifts $c^{\dagger}_{b,\bfk} \to e^{-i \bfr_b \cdot \bfk } c^{\dagger}_{b,\bfk}$, with different offsets $\bfr_b$ for each sublattice $b$, so that the transformed Bloch functions are invariant under $\bfk \to \bfk + \bfG$: the transformed curvature and metric are then periodic as well. The resulting Hamiltonian is, of course, gauge-inequivalent to the original one, which can be seen from the fact that the curvature itself changes: under transformations 
\begin{equation}
\label{eq_not_gauge}
u^\alpha_a(\bfk) \to \widetilde{u}^\alpha_b(\bfk) = e^{i \bfr_b \cdot \bfk} u^\alpha_b(\bfk)
\end{equation}
for $b= 1, \ldots, \calN$, the Berry curvature at $\bfk$ changes by  
\begin{equation}
\widetilde{B}_\alpha (\bfk) - B_\alpha (\bfk)= \sum_{b = 1}^{\calN} r_{b,y} \frac{\partial}{\partial k_x} |u^\alpha_b(\bfk)|^2 -r_{b,x} \frac{\partial}{\partial k_y} |u^\alpha_b(\bfk)|^2.
\end{equation}
Because this is a sum of total derivatives, the surface integrals of $\widetilde{B}_\alpha (\bfk)$ and $B_\alpha (\bfk)$ yield the same Chern number. The difference itself, however, only vanishes when $\bfr_b$ is the same for all $b$, which is the gauge transformation (\ref{eq_bloch_gauge}).

Phase shifts of the form (\ref{eq_not_gauge}) were employed in recent publications\cite{Wu:2012do,Lee:2013bz} to obtain band Hamiltonians that were periodic in $\bfk$. This was described as a ``gauge transformation'' in these references, but as we've noted, the only gauge symmetry of the Hamiltonian is with respect to \emph{bands} (i.e. $U(1)$ rotations in the eigenbasis $\gamma^{\alpha \dagger}_{\bfk} \to e^{-i \phi_\alpha(\bfk)} \gamma^{\alpha \dagger}_{\bfk}$). For the transformations made in Supplementary Ref.~\onlinecite{Wu:2012do}, the difference in curvature fluctuations is substantial, as shown in Supplementary Fig.~\ref{fig-zoo} for the Haldane, kagom\'e lattice and ruby lattice models. Note that each panel of this figure shows the curvature distribution for parameters which minimize $\sigma_{B}$ as computed with each panel's respective Hamiltonian; i.e. they depict the closest one may get to uniform band geometry in the parameter space of the Hamiltonian considered.

For completeness, we note that unlike these single-particle properties, the many-body gap is invariant under the generalized phase shift (\ref{eq_not_gauge}), because single-particle density operators $\olrho_\bfk$ are left unchanged by the transformation. As an example, in Supplementary Fig.~\ref{fig-zoo26} we reproduce the results shown in Fig.~26 of Supplementary Ref.~\onlinecite{Wu:2012do}, along with the fluctuations in the canonically defined Berry curvature for the same system. The latter are lower than in the transformed Hamiltonian used in that reference, meaning that the negative correlation between $\sigma_B$ and the gap is stronger than depicted there: introducing phase shifts by hand distorts the Berry curvature distribution to a degree which significantly affects the conclusions one may draw from that data.

We have shown that band geometry may be defined for non-periodic band Hamiltonians; we now argue that the non-periodic basis used in this paper is in fact the one measured by any (direct or indirect) experimental probe, and hence is the only one which should be regarded as physical. The feature possessed by Bloch bands which is absent from the general theory of Berry phases is the fact that the parameter space in question is defined via the Fourier transform of the kinematic setting of the physical system. Because this is a \emph{global} transform, in doing the Fourier sum 
\begin{equation}
\label{eq_tb_fourier}
|\bfk,b\rangle = \frac{1}{\sqrt{N_c}} \sum_{\bfR} e^{i\bfk \cdot(\bfR + \bfd_b)} |\bfR, b\rangle,
\end{equation}
we have already implicitly chosen a basis for the band Hilbert space at each $\bfk$. For example, the  the transformations made in Supplementary Refs. \onlinecite{Wu:2012do,Lee:2013bz} correspond to defining modified tight-binding states on the reciprocal lattice via
\begin{equation}
\widetilde{\chi}_b (\bfk) = \frac{1}{\sqrt{N_c}} \sum_{\bfR} e^{i\bfk \cdot \bfR} \chi_b(\bfr - \bfR - \bfd_b)
\end{equation}
where $\chi_b(\bfr - (\bfR + \bfd_b)) = \langle \bfr | \bfR b \rangle$; this is manifestly invariant under $\bfk \to \bfk+ \bfG$, but in doing so the position operator is no longer consistently defined for orbitals with different offsets $\bfd_b$: the transformation is equivalent to shifting the orbitals of the crystal basis to the origins of various lattice cells. As an obviously apparent consequence, the crystal symmetry of the transformed Berry curvatures in the second row of Supplementary Fig.~\ref{fig-zoo} are unphysically broken. Because experiments probe position-space quantities, we conclude that the basis defined by \eqref{eq_tb_fourier} and used in this paper is the canonical, physically relevant quantity.


\section{Relation between the trace and determinant inequalities}
\setcounter{equation}{11}
\label{sec-ineqs}
The inequalities 
\begin{align}
\label{eq_trineq}
\Tr g^\alpha (\bfk) &\geq |B_\alpha(\bfk)|; \\
\label{eq_detineq}
\det g^\alpha(\bfk) &\geq B_\alpha(\bfk)^2/4,
\end{align}
were proved in Supplementary Ref.~\onlinecite{Roy:2012vo}. Assuming the quantum metric is nondegenerate, it may be factored as
\begin{equation}
g^\alpha = \sqrt{g^\alpha} \begin{pmatrix} \tilde{g}_{11} & \tilde{g}_{12} \\ \tilde{g}_{12} & \tilde{g}_{22} \end{pmatrix}
\end{equation}
where the scalar $\sqrt{g^\alpha} = (\det g^\alpha)^{1/2}$ and the second factor is a unimodular matrix. Using the standard inequality between arithmetic and geometric means,
\begin{equation}
\Tr g^\alpha = \sqrt{g^\alpha}(  \tilde{g}_{11}+ \tilde{g}_{22}) \geq 2\sqrt{g^\alpha}\sqrt{  \tilde{g}_{11}\tilde{g}_{22}}.
\end{equation}
By unimodularity $\sqrt{  \tilde{g}_{11}\tilde{g}_{22}} = \sqrt{ 1+ \tilde{g}_{12}^2} \geq1$, so 
\begin{equation}
\left(\Tr g^\alpha(\bfk) \right)^2 \geq 4\det g^\alpha(\bfk),
\end{equation}
with equality if and only if $g^\alpha$ is proportional to the identity matrix. Because the Berry curvature is algebraically independent of the components of the quantum metric, we can conclude that saturation of trace inequality (\ref{eq_trineq}) implies saturation of the determinant inequality (\ref{eq_detineq}). In physical terms, a constant curvature and metric that saturate the determinant inequality imply the closure of a modified Girvin-MacDonald-Platzman (GMP) algebra; saturation of the trace inequality then corresponds to the stronger condition that the algebra of band projected density operators is not only isomorphic to the GMP algebra, but identical to it.


\section{Two-band models}
\label{sec-twoband}
\setcounter{equation}{16}
By adding suitably chosen exponentially-localized couplings, any two-band Hamiltonian may be brought to a band-flattened form parameterized by a unit 3-vector $H_\text{flat}(\bfk) = - \hatn(\bfk) \cdot \sigma$, where $\sigma = (\sigma_1, \sigma_2, \sigma_3)$ are the three Pauli matrices. The eigenvectors of the original and flattened Hamiltonians are identical (by construction) and can be obtained analytically:
\begin{equation}
u^\pm(\bfk) = \frac{1}{\sqrt{2(1 \mp \hatn_3)}} \begin{pmatrix} \hatn_3 \mp 1 \\  \hatn_1 + i  \hatn_2 \end{pmatrix}.
\end{equation}
Using this expression gives the Berry curvature and quantum metric of the lower band as
\begin{align}
B &= \tfrac{1}{2} \hatn \cdot \partial_x \hatn \times \partial_y \hatn; \\
g_{\mu \nu}&= \tfrac{1}{4} \partial_\mu \hatn \cdot \partial_\nu \hatn,
\end{align}
since $\hatn \cdot \partial_\mu \hatn = 0$. Using this and standard identities relating multiple dot and cross products, it follows that
\begin{equation}
g_{xx}(\bfk) g_{yy}(\bfk) - g_{xy}(\bfk)^2 -\frac{1}{4} B(\bfk)^2 =0:
\end{equation}
the determinant condition $\det g^\alpha(\bfk) - B_\alpha(\bfk)^2/4 = 0$ is necessarily satisfied for any two-band model. As our results with the kagom\'e and ruby lattice models show, this ceases to be the case for models having more than two bands.


\section{Scaling of gaps with the number of bands}
\label{sec-crossmodel}
\setcounter{equation}{20}

Fractional Chern insulators (FCIs) exhibit states which may be thought of as discretizations of continuum fractional quantum Hall (FQH) states, in that they have identical topological and long-wavelength properties (see, e.g., Supplementary Ref. \onlinecite{Regnault:2011bu}). To that end, consider a continuum position-space wavefunction $\psi(\bfr)$ which is discretized to set of values $\psi_{b}$ in a tight-binding model with $\calN$ sites per unit cell. The normalization conditions on the corresponding Bloch functions are
\begin{equation}
\int_{\text{UC}} d\bfr \, |u_{\bfk} (\bfr)|^{2} = \sum_{b=1}^{\calN} |u_{\bfk,b}|^{2} = 1,
\end{equation}
for any $\bfk$, where the integral is taken over a unit cell of the lattice. In the limit of large $\calN$, we may assume that $u_{\bfk}(\bfr)$ at $\bfr = \bfd_{b}$ is proportional to the discretized value $u_{\bfk,b}$. Approximating the continuum normalization integral by a sum introduces a factor of $1/\calN$ from the integration measure, which is compensated by the scaling
\begin{equation}
\label{eq_bloch_norm}
u_{\bfk}(\bfd_{b}) \sim \sqrt{\calN} u_{\bfk,b}.
\end{equation}

Now consider the matrix elements of the delta-function interaction employed in the main text to stabilize the bosonic Laughlin state. In the continuum, these are
\begin{align}
\langle \bfk_{3} \bfk_{4}| \widehat{V} | \bfk_{1} \bfk_{2} \rangle &= V \int d \bfr \, u_{\bfk_{3}}^{\ast}(\bfr) u_{\bfk_{4}}^{\ast}(\bfr) u_{\bfk_{1}}(\bfr) u_{\bfk_{2}}(\bfr)  \nonumber \\
& \sim \frac{V}{\calN} \sum_{b} u_{\bfk_{3}}^{\ast}(\bfd_{b}) u_{\bfk_{4}}^{\ast}(\bfd_{b}) u_{\bfk_{1}}(\bfd_{b}) u_{\bfk_{2}}(\bfd_{b}), \nonumber
\end{align}
times a momentum-conserving $\delta$-function. Comparing this with an on-site interaction in the discretized model
\begin{equation}
V_{\text{disc}} \sum_{b} u_{\bfk_{3},b}^{\ast} u_{\bfk_{4},b}^{\ast} u_{\bfk_{1},b} u_{\bfk_{2},b} \nonumber
\end{equation}
and using \eqref{eq_bloch_norm} shows that the discretized interaction strength should be scaled as 
\begin{equation}
\label{eq_d2_scaling}
V_{\text{disc}} \sim V \calN \qquad \text{($\delta$-function).}
\end{equation}
This means that, given gaps which have been obtained for the FCI Laughlin state in two lattice models $A,B$, the quantities which should be compared are $\calN_{A}\Delta_{A}$ and $\calN_{B}\Delta_{B}$ (assuming that the single-particle dispersion has been flattened; i.e. that the gap is only set by the interaction term.) Similar considerations show that for the three-body delta-function interaction used to stabilize the Moore-Read state, the leading scaling should be $V_{\text{disc}} \sim V \calN^{2}$.

We note that the above argument is not as simple as it may appear: we're implicitly assuming $\psi(\bfr)$ has the character of a low-lying eigenstate --- more specifically, that it has support over almost all tight-binding orbitals. This is the case when $u_{\bfk,b}$ is a randomly chosen vector from the $\calN$-dimensional band Hilbert space, but one can easily construct counterexamples violating this assumption: for example, let lattice model $B$ (with $\calN_{B} > \calN_{A}$) have a block-diagonal kinetic term, the first block of which is identical to the kinetic term of model $A$ and the second block of which describes additional, trivial ``spectator'' orbitals with very high occupation energies. The low-lying bands $u^{B}_{\bfk,b}$ will have vanishing amplitude on these spectator orbitals, and will numerically be identical to the corresponding eigenfunctions $u^{A}_{\bfk,b}$ of model $A$, with zeros appended. In this scenario, the eigenfunction scaling assumption \eqref{eq_bloch_norm} is violated, and the interaction strength (and gap size) would \emph{not} scale with the number of spectator orbitals.

The scaling argument can be placed on a more rigorous footing in the context of the Hofstadter model\cite{Hofstadter:1976js} in the limit of small flux per plaquette $\phi =1/N$, which offers a sequence of lattice models (with $N$ bands) which converges to the continuum FQHE in the $N \to \infty$ limit. In this case, the scalings \eqref{eq_bloch_norm}, \eqref{eq_d2_scaling} are numerically observed to be obeyed to high accuracy\cite{hof-temporary}. For the case of the FCI models studied in this work, we know the ground states of different models lie in the same universality class as the continuum FQHE state (as can be determined by topological order, done here by entanglement spectrum counting), but we do not expect, e.g., the ruby lattice model ground state to be an interpolation of the kagom\'e lattice model ground state. The fact that we observe the scaled gaps to be so close to each other provides an \emph{a posteriori} confirmation of this argument's validity even for the relatively small values of $\calN = 2,3,6$.


\section{Expansion of band-geometric quantities in parameter space}
\label{sec-gradients}
\setcounter{equation}{23}

The Berry curvature and quantum metric are defined in terms of the derivatives of a wavefunction over some parameter manifold. For both numerical and theoretical purposes, it becomes convenient to recast these expressions in terms of derivatives of the Hamiltonian itself, rather than its eigenfunctions, since ordinarily only the former is known analytically. This, of course, is not a new observation \cite{Berry:1984ka}. 

In order to avoid overall phase ambiguities and the necessity of using multiple charts to cover the Brillouin zone (BZ) in topologically non-trivial situations, it's preferable to work with the occupied band projector $P_\alpha = |\bfk,\alpha \rangle \langle \bfk, \alpha|$, instead of the eigenfunction $|\bfk,\alpha \rangle$ itself. In these terms, the Feynman-Hellman theorems are
\begin{align}
\label{eq_de_def}
\partial_\lambda E_\alpha &= \Tr P_\alpha \, \partial_\lambda H ; \\
\label{eq_dp_def}
\partial_\lambda P_\alpha &= R_\alpha \, \partial_\lambda H \, P_\alpha + P_\alpha\, \partial_\lambda H \, R_\alpha,
\end{align}
where the projected resolvent operator $R_\alpha$ is
\begin{equation}
R_\alpha = \sum_{\beta \neq \alpha} \frac{|\beta \rangle \langle \beta |}{E_\alpha - E_\beta} =  \frac{1-P_\alpha}{E_\alpha - H}.
\end{equation}
Equations (\ref{eq_de_def}) and (\ref{eq_dp_def}) hold for any parameter $\lambda$ upon which the Hamiltonian smoothly depends, and they are valid for arbitrary values of $\lambda$ and hence may be further differentiated. 

Assuming no additional degeneracies, $(E_\alpha - H)$ may be inverted in the subspace orthogonal to $|\bfk, \alpha \rangle$, and
\begin{align}
\label{eq_dr_def}
\partial_\lambda R_\alpha &=  -R_\alpha^2 \, \partial_\lambda H \, P_\alpha  -P_\alpha \, \partial_\lambda H \, R_\alpha^2 - R_\alpha \partial_\lambda (E_\alpha - H )\, R_\alpha.
\end{align}
The relations (\ref{eq_de_def}), (\ref{eq_dp_def}) and (\ref{eq_dr_def}) then form a closed system and may be iterated to any order to develop the Taylor expansion of band-geometric quantities.

For example, the Berry curvature and quantum metric for a single occupied Chern band are\cite{Avron:1985fo,CamposVenuti:2007il}
\begin{align}
\label{eq_b_curvature}                             
B_\alpha &= 2\, \Imag \Tr P_\alpha \, \partial_y H \, R^2_\alpha \, \partial_x H , \\
\label{eq_q_metric}  
g^\alpha_{\mu\nu} &= \Real \Tr  P_\alpha \, \partial_\mu H \, R^2_\alpha \, \partial_\nu H.
\end{align}
The derivative of the Berry curvature with respect to a coupling $\lambda$ follows from a straightforward computation as
\begin{align}
\partial_\lambda B_\alpha &= 2 \, \Imag \Tr P_\alpha \left\{ -2 (\partial_\lambda E_\alpha) \partial_y H \, R^3_\alpha \, \partial_x H + \partial_y H \, R^2_\alpha \, \partial_{x,\lambda}H + \partial_{y, \lambda} H \, R^2_\alpha \, \partial_x H \nonumber \right. \\
{}& \qquad \qquad + \partial_y H \, R_\alpha ( \partial_\lambda H \,R_\alpha + R_\alpha \, \partial_\lambda H ) R_\alpha \, \partial_x H \nonumber \\
\label{eq_curvature_gradient} & \qquad \qquad + \left. \left[ \left( \partial_\lambda H \, R_\alpha \, \partial_y H \, R^2_\alpha \, \partial_x H + (\partial_x E_\alpha)  \partial_\lambda H \, R^3_\alpha \, \partial_y H \, \right)  - ( x \leftrightarrow y) \vphantom{\langle \partial_y H\rangle}  \right] \right\}.
\end{align}
Other quantities such as the Hessian $\partial_\lambda \partial_{\lambda'} B_\alpha$ may be calculated in a similar manner, although the algebra rapidly becomes tedious. These may, in turn, be used to find the variation in Brillouin zone-averaged quantities with respect to couplings in the Hamiltonian; for example, denoting by $\langle \cdots \rangle$ the Brillouin zone average,\begin{align}
c_1 &= \frac{A_{BZ}}{2\pi} \langle B \rangle,\\
\sigma_B &= \frac{A_{BZ}}{2\pi} \sqrt{ \langle B^2 \rangle -  \langle B \rangle^2}, \\
\label{eq_rms_curvature_gradient} \text{and so } \partial_\lambda \sigma_B &=  \frac{(A_{BZ}/ 2\pi)^2}{ \sigma_B} \langle B \partial_\lambda B \rangle.
\end{align}
Here we've made use of the fact that $\partial_\lambda \langle B \rangle = 0$ while we remain in the same topological phase, which can easily be checked during numerical computations. Using (\ref{eq_curvature_gradient}) in (\ref{eq_rms_curvature_gradient}) then permits us to rapidly perform a steepest-descent minimization of Berry curvature fluctuations, even in a high-dimensional parameter space.

\clearpage
\section*{Supplementary References}
\bibliographystyle{naturemag}

\end{document}